\documentstyle [11pt,epsf] {article}

\begin{document}

\begin{center}
\vspace{2cm}
\LARGE
Radio Jets in Galaxies  with Actively Accreting Black Holes:\\
new insights from the SDSS
\\                                                     
\vspace{1cm} 
\large
Guinevere Kauffmann$^1$, Timothy M. Heckman$^2$, Philip N.  Best$^3$\\                   
\vspace{0.5cm}
\small
\noindent
{\em $^1$Max-Planck Institut f\"{u}r Astrophysik, D-85748 Garching, Germany} \\
{\em $^2$Department of Physics and Astronomy, Johns Hopkins University, Baltimore, MD 21218}\\
{\em $^3$Institute for Astronomy, Royal Observatory Edinburgh,
Blackford Hill,  Edinburgh EH9 3HJ, Scotland, UK}\\

\vspace{1.5cm}

\begin {abstract}

In the local Universe, the majority of radio-loud AGN are found in 
massive elliptical galaxies with old stellar populations and
weak or undetected emission lines. 
At high redshifts, however, almost all known radio AGN have strong emission lines. 
This paper focuses on a subset of radio AGN with emission lines (EL-RAGN) 
selected from the Sloan Digital Sky Survey. We explore the hypothesis  
that these objects are local analogs of powerful high-redshift radio galaxies. 
The probability 
for a nearby  radio AGN to have  emission lines
is a strongly decreasing function of galaxy mass and  velocity dispersion and
an increasing function of radio luminosity above  $10^{25}$ W Hz$^{-1}$.
Emission line and radio luminosities are 
correlated, but with  large 
dispersion. At a given radio power, radio galaxies  with small    
black holes  have higher [OIII] luminosities 
(which we interpret as  higher accretion rates)
than radio galaxies with big black holes. However, if we scale the emission line 
and radio luminosities by the black hole mass,  we find a correlation between 
normalized radio power and accretion rate in Eddington units that is independent
of black hole mass. There is also a clear correlation between
normalized radio power and the age of the stellar population
in the galaxy.  Present-day EL-RAGN with the 
highest normalized radio powers are confined to 
galaxies with small black holes. 
High-redshift, high radio-luminosity  AGN would be explained if big black
holes were similarly active at earlier cosmic epochs. 

To investigate why only a small fraction of emission
line AGN become radio loud, we create matched samples of radio-loud
and radio-quiet AGN and compare their host galaxy properties and environments.
The  main difference  
lies in their environments; our local density estimates
are a factor 2 larger around the radio-loud AGN. We propose a                
scenario in which radio-loud AGN with emission lines are located in galaxies
where accretion of both cold and hot gas can occur simultaneously.
At the present day, these conditions are only satisfied for low mass galaxies in
dense environments, but they are likely to apply to most            
galaxies with  massive black holes at higher redshifts.   

\end {abstract}
\end{center}
\vspace{1.5cm}
\normalsize
\pagebreak

\section {Introduction}

Over the last few years, there has been growing realization 
that the powerful relativistic jets associated with  radio
galaxies may play a very important role in regulating
the star formation histories of the most massive galaxies in the
Universe. X-ray studies of groups and clusters of galaxies with
{\em Chandra} and {\em XMM-Newton} have shown that these jets interact
strongly with their environment, blowing clear cavities or ``bubbles''
in the surrounding X-ray emitting gas (B\"ohringer et al. 1993; Fabian et al.
2000,2003,2005; Churazov et al. 2001; Birzan et al. 2004; Forman et al. 2005).  
Studies of the demographics of nearby  radio galaxies show that
the jet phenomenon is  primarily associated with the most massive
galaxies and black holes in the local Universe (Best et al 2005b; hereafter B05b).
Radio galaxies are also found to be preferentially located in central
group and cluster galaxies when compared to other galaxies 
of the same mass (Best et al 2007). Gas cooling rates are expected to 
be the highest in these systems,  but it can be shown that 
the integrated energy input by 
radio jets is likely to be sufficient to shut off the cooling and 
star formation in these  massive galaxies, leading naturally to the
observed exponential cutoff at the bright end of the galaxy  
luminosity function (B05b; Best et al 2006; Croton et al 2006; Allen et al 2006).

It has long been known that radio galaxies can be divided into
two major classes (Fanaroff \& Riley 1974). In Class I (FRI) sources,
the radio emission peaks near the center of the galaxy and the 
emission from the jets fades with distance from the center. 
Class II (FRII) sources have ``edge-brightened'' radio lobes.
Class I sources dominate at low radio power and low redshifts, while
high power radio galaxies with 178MHz radio powers greater than $10^{27}$
W Hz$^{-1}$ are almost exclusively FRII systems. 
Another way of classifying radio galaxies is according to whether they
have strong high-excitation narrow-line emission. The objects lacking
emission lines are sometimes referred to as low-excitation radio
galaxies (Hardcastle, Evans \& Croston 2006).  The majority
of low-power, FRI radio galaxies are low excitation systems. Conversely, 
most powerful, high-redshift FRII radio galaxies have strong emission lines.      

In the local Universe, the majority of radio galaxies are  
relatively low power FRI radio sources. They are found in typical elliptical
galaxies with very little ongoing star formation and weak
emission lines (Ledlow \& Owen 1995; Govoni et al 2000; B05b).
If optical emission lines are present, they  generally have s LINER-type  line ratios.
X-ray studies of low-excitation radio galaxies are consistent with
a model in which the active nuclei of these objects are not radiatively
efficient in any waveband (Hardcastle, Evans \& Croston 2006).  
There is almost never any evidence of a heavily absorbed nuclear
X-ray component, which suggests that a  classical accretion disk
and obscuring torus is not  present in these systems.
Recently Allen et al (2006) have shown that a tight correlation exists
between the Bondi accretion rates calculated from the observed gas
temperature and density profiles and estimated black hole masses, and
the power emerging from these systems in relativistic jets.
This result suggests that the jets are being powered by accretion from
hot gas halos surrounding the galaxy.

In the nearby Universe, 
powerful FRII radio galaxies have much lower volume densities compared to
the FRIs, but the FRII space density appears to increase  very strongly as a 
function of redshift (Laing, Riley \& Longair 1983; Dunlop \& Peacock 1990).    
It has  been established that powerful FRIIs exhibit a strong 
correlation between their radio and  their emission line
luminosities (Baum, Zirbel, O'Dea 1995). This suggests that 
both the optical and the radio emission are linked to the same physical
process. It is also generally accepted that powerful FRII radio galaxies   
and radio-loud quasars are likely to be the same objects, observed
at different viewing angles to the axis of the jet (Barthel 1989).
As such, the black holes in powerful FRIIs are undergoing {\em radiatively 
efficient} accretion and the optical line emission is likely powered 
by gas from an accretion disk. 

It should be noted that although there is a gross connection between 
Fanaroff-Riley class and the presence or absence of emission lines,
the relation is certainly not one-to-one. 
A significant number of reasonably powerful FRII radio galaxies  
are in fact low excitation systems (Hine \& Longair 1979; Laing et al 1994;
Hardcastle et al 2006). This indicates that the transition between FRI and FRII radio
morphology and that between radiatively efficient and inefficient
accretion are different phenomena, even though both occur at similar
radio luminosities. The physical origins of this
difference are not understood.

Understanding the circumstances under which an
accreting black hole is able to produce a radio-jet is extremely important
if we wish to understand how AGN  feedback can  affect galaxies at different
redshifts. In current theoretical models, feedback from radio AGN is
only assumed to play an important role  for massive black holes located
at the centers of dark matter halos that contain a substantial reservoir
of hot X-ray emitting gas (Croton et al 2006; Bower et al 2006; 
Sijacki et al 2007).
At high redshifts, massive X-ray emitting halos are rare and it has been postulated
that large-scale outflows from quasars become more important in regulating the
growth of the most massive galaxies (Di Matteo, Springel \& Hernquist 2005; 
Hopkins et al 2005). The FRII radio galaxy phenomenon has 
largely been neglected up to now.
       
Ideally, one would like to carry out a controlled comparison of 
different classes of radio-loud and radio-quiet AGN in order to 
address some of the issues discussed above. It is important that  different 
AGN classes be compared at the same redshift, so that evolutionary effects
do not complicate the analysis. Recently,  B05b compared the
host galaxy properties of radio-loud AGN with those of optically-
selected narrow-emission line AGN. Both samples were drawn from the main
spectroscopic sample of the Sloan Digital Sky Survey. As mentioned previously, 
the sample of radio galaxies analyzed by B05b was dominated
by very massive early-type galaxies with weak or undetectable   
optical emission lines. 
A minority of radio galaxies with emission lines 
can be  found in this sample, however, 
and it is these objects which are the focus
of the analysis in the present paper. We use our sample of radio galaxies
with emission lines to address the following questions:

\begin {enumerate}
\item Is there a correlation between radio luminosity and emission line
luminosity for these low redshift systems? If so, how does this compare
with results in the literature, which have generally included radio
galaxies at higher redshifts than those in our sample?
\item How do the host galaxies of radio AGN with emission lines compare
with those of radio galaxies without emission lines? 
\item How do the host galaxy  and environments of radio AGN               
with emission lines compare with those of radio-quiet emission line AGN? 
\item Do these comparisons shed any light on why some AGN develop radio jets?
Can we make a link between the optical and radio AGN phenomena and
different gas accretion modes as has been proposed in a recent
paper by Hardcastle, Evans \& Croston (2007)? 
\item Can we find any evidence that feedback from the radio AGN influences 
the star formation histories and/or structural properties of the host
galaxies of these systems?
\end {enumerate}

Our paper is organized as follows.  In section 2, we review how the radio
galaxies in our sample were selected  and describe how we differentiate
between star-forming galaxies and radio AGN with emission lines.
In section 3, we   study the demographics of the radio
galaxies with emission lines and the  correlations between their radio and emission
line properties.  In section 4, we compare the
host galaxy properties and environments of radio-loud AGN  with emission
lines to those of matched samples of radio-quiet AGN. In sections 5 and 6, we
compare the host galaxy properties and environments of radio-loud
AGN to those of matched samples of  galaxies that are selected
without regard to their AGN properties  and we ask 
whether there is any evidence that the presence or
absence of a radio jet has  an influence on
the recent star formation history of the galaxy. Our conclusions
are presented in section 7.

\section {The Sample}

\subsection {Selection of radio-loud AGN}

Best et al (2005a; hereafter B05a) identified radio-emitting galaxies within the
main spectroscopic sample of the SDSS data release 2 (DR2) by
cross-comparing these galaxies with a combination of the 
National Radio Astronomy Observatory (NRAO) Very Large Array (VLA)
Sky Survey (NVSS; Condon et al. 1998) and the Faint Images of
the Radio Sky at Twenty centimeters (FIRST) survey (Becker et al. 1995).
The combination of these two radio surveys allowed a 
radio sample to be constructed  that was both reasonably complete ($\sim$95\%)
and highly reliable (i.e. it is estimated that $\sim$99\% of the sources
in the catalogue are genuine radio galaxies rather than
false matches). In this paper, we make use of
an updated catalogue of radio galaxies based on the SDSS data release 4 (DR4).

Once the radio and optical catalogues have been cross-compared, 
the next task is to separate radio AGN from galaxies in which the
radio emission comes primarily from star-forming regions in the galaxy.
The ideal would be to  identify radio 
AGN as those galaxies with radio flux densities that are significantly
in excess of that  predicted using the  far-infrared/radio correlation
for star-forming galaxies (e.g. Yun, Reddy \& Condon 2001).
However, sufficiently deep IR data were not available for this
to be a viable method for the SDSS sample, so B05a separated
the two populations using the location of the galaxy in the plane
of 4000 \AA\ break strength (D$_n$4000) versus radio luminosity per
unit stellar mass. The division between star forming
galaxies and AGN  was motivated using  Bruzual \& Charlot (2003)
stellar population synthesis predictions for how galaxies
with different star formation histories populate this plane.
The nominal dividing line was chosen to lie 0.225 in D$_n$(4000) above
the curve defined by galaxies undergoing exponentially declining star
formation with a time constant $\tau =3$ Gyr (see Fig. 9 of B05a).     

One  problem with the B05a criterion is that  all galaxies with D$_n$(4000)
values less than 1.3 will be excluded from the radio AGN sample, 
regardless of their radio luminosity.
In this paper, we  present an alternative way of separating radio AGN from
star-forming galaxies. The method is based on the correlation between 
radio luminosity  and ``corrected'' H$\alpha$
luminosity plotted in Figure 1 . The H$\alpha$ luminosty has been corrected
for dust extinction using the measured Balmer decrement
\footnote {Note that in this paper, we correct emission lines such as H$\alpha$ and   [OIII] 
for extinction using the Balmer decrement,  assuming a
dust-free value of $H\alpha$/H$\beta$ of 2.86 and an extinction
curve of the form $\lambda ^{-0.7}$ (Charlot \& Fall 2000). }, 
and for fibre
aperture effects. To correct for the aperture effect, we simply scale the
measured H$\alpha$ luminosity by the ratio of the galaxy's Petrosian  
magnitude to its fibre magnitude measured in the $r$-band. 
This is only a crude correction, but as can be seen from the left-hand panel 
of Figure 1, galaxies classified as   
star-forming using the standard [OIII]/H$\beta$ versus [NII]/H$\alpha$
Baldwin, Phillips \& Terlevich (1981; hereafter BPT) diagram,  define a rather
tight correlation in the $\log P/ \log L(H\alpha)$ plane \footnote {See Kauffmann
et al (2003b) for details about how galaxies are classified as
star-forming or as AGN using the BPT diagnostic diagrams.}. 
Almost all the outliers have NVSS fluxes larger than 5 mJy and
are probably radio AGN. We use this tight correlation to define
an  AGN/SF separation line in this diagram (the dashed line). 

We can  test the robustness of our method by checking whether              
there is   internal consistency between dividing lines that are
based on different star formation indicators.
 Figure 2 shows radio AGN and star-forming galaxies  
 selected using  
the $\log P$--$\log L(H\alpha)$ separator   defined in Figure 1.
We plot the location of these objects in the plane of D$_n$4000
versus $\log P/M_*$.
As can be seen,  
radio AGN and star-forming galaxies
segregate extremely cleanly in D$_n$(4000)--$\log P/M_*$ space.
This proves that both diagrams can  be used interchangeably.
The dividing line in D$_n$(4000) versus $\log P/M_*$ space that we
find is slightly different to the cut proposed by B05a, which
is shown as a dashed line on the plot. Nevertheless 
our test indicates that 
the method is {\em in principle} quite clean.

\begin{figure}
\centerline{
\epsfxsize=15cm \epsfbox{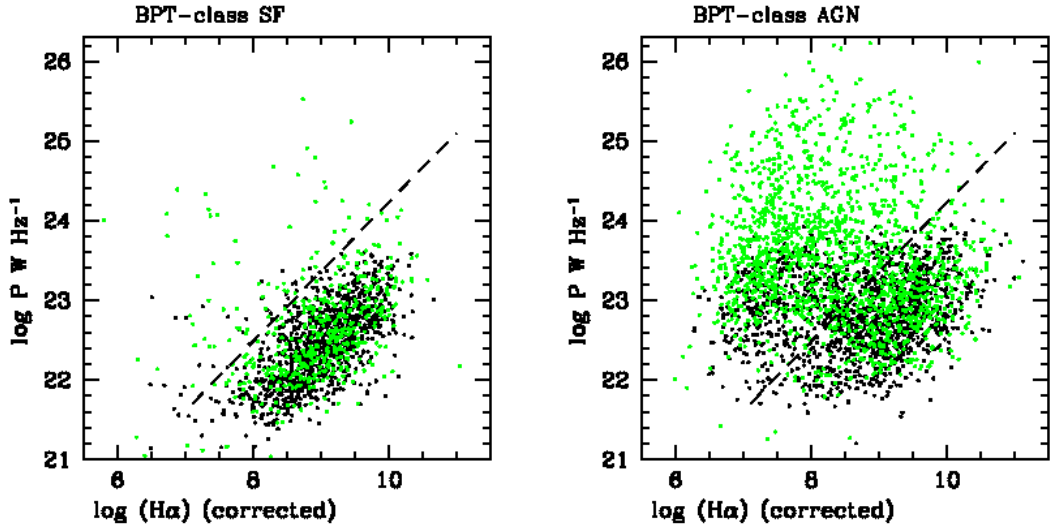}
}
\caption{\label{fig1}
\small
Separation of radio AGN and star forming-galaxies in              
$\log P$-- $\log L(H\alpha)$ space. The left panel shows galaxies
classified as star-forming from their
emission line ratios using the BPT diagram diagnostic, while the right panel
shows galaxies classified as optical AGN.
Black points show galaxies with  S(NVSS) $ > 2.5$ mJy. Galaxies with
 S(NVSS)$ > 5$ mJy have been overplotted using green. The dashed line is 
our proposed separation line for distinguishing radio AGN from
galaxies where the radio emission is coming from star-forming regions.
Note that L(H$\alpha$) is given in solar units and the NVSS fluxes
are measured at 1.4 GHz. }
\end {figure}
\normalsize

\begin{figure}
\centerline{
\epsfxsize=15cm \epsfbox{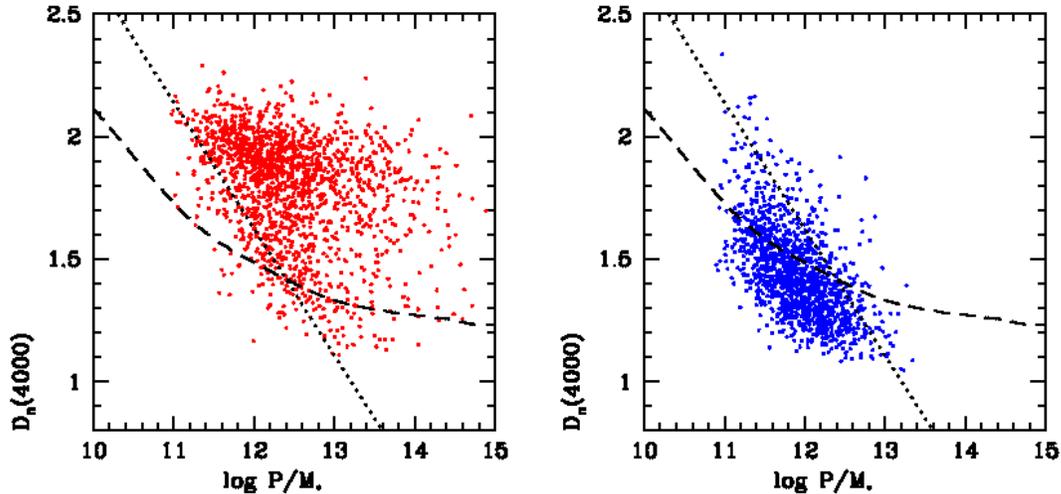}
}
\caption{\label{fig1}
\small
A consistency check of the two methods (see text). 
We plot D$_n$(4000) as a function of
$\log P/M_*$ for  radio AGN (left) and for star-forming galaxies (right) defined   
using the  $\log P$--$\log L(H\alpha)$ separation line 
plotted  in Figure 1. The dashed line shows the separator defined
in B05a. The dotted line has been included to guide the eye.}
\end {figure}
\normalsize

\subsection {Galaxy properties from the optical spectra}
A variety of physical properties have been derived for galaxies
in the spectroscopic database via stellar population synthesis fitting
and are publically available \footnote{http://www.mpa-garching.mpg.de/SDSS}.
The stellar continuum of each galaxy is modelled as a sum of template
spectra generated from population synthesis models (Tremonti et al 2004).
These fits also lead to measures of  stellar mass-to-light ratios,
star formation histories and mean stellar ages (Kauffmann et al 2003a).
After subtracting the stellar continuum, emission line fluxes can be
accurately measured, allowing studies of the star formation rates 
(Brinchmann et al 2004) and AGN properties (Kauffmann et al 2003b;
Heckman et al 2004) of the galaxies in the sample.

\section{Properties of  radio galaxies with emission lines}

\subsection {Demographics}
In Figure 3, we show how the fraction of  radio galaxies with emission lines  
depends on galaxy properties such as stellar mass, velocity dispersion
and mean stellar age, as well as  on  radio luminosity.
A radio galaxy is defined to have emission lines if the  
four emission lines  H$\alpha$, 
H$\beta$, [OIII] and [NII] are all detected with signal-to-noise greater than
3.

In past work, we have often used the 4000 \AA\  break index D$_n$(4000) as
an indicator of stellar age. As can be seen in Figure 3, the fraction
of radio galaxies that have emission lines is a strong function of
D$_n$(4000): almost all radio galaxies with D$_n$(4000) less than 1.6 have
emission lines, and the fraction falls to below a third for D$_n$(4000) $ >
1.8$. This result suggests that the presence of emission lines and the
presence of a young stellar population in the galaxy are strongly
correlated. However, low values of D$_n$(4000) would also result if the
stellar light is diluted by non-thermal AGN emission, and there has
been ongoing controversy as to whether the UV continuum light from
powerful radio galaxies arises from the AGN itself (either direct AGN
light or light that is scattered into the line-of-sight), from young
stars or from nebular continuum (see Tadhunter et al 2002 for a
discussion). Therefore, to avoid any ambiguity in the interpretation
of our results, we also show in Figure 3 the correlation between radio
power and the Balmer absorption line Lick index H$\delta_A$
(Worthey \& Ottaviani 1997). Non-thermal light from the AGN would act
to dilute the absorption lines and lower the value of H$\delta_A$. Any
trend towards higher values of H$\delta_A$ should be regarded as a
definitive indication of the presence of young stars in the host
galaxy. The top-right panel of Figure 3 demonstrates that the fraction
of radio galaxies that have emission lines increases strongly with 
increasing H$\delta_A$, confirming the correlation between emission line 
activity and a young stellar population found in the D$_n$(4000) analysis.

The fraction of emission line radio galaxies also decreases as a 
function of the stellar mass and velocity dispersion of the galaxy.  
B05b showed that overall the fraction of radio loud galaxies is a strongly
{\em increasing} function of both stellar and black hole mass. 
The results in the two bottom left  panels                             
of Figure 3 indicate that the fraction of all radio loud AGN that  
have  emission lines has the opposite dependence on stellar mass. As a result, 
radio AGN with emission lines  are a minority of all the radio-loud AGN in our sample (1684
out of 5714 objects with NVSS fluxes larger than 2.5 mJy).

Finally, in the bottom right panel of Figure 3, we plot the fraction of
emission line radio AGN as a function of the radio luminosity.
The fraction remains constant at a value of around 0.4 up to
a radio luminosity of $10^{25}$ W Hz$^{-1}$ and then appears to rise.
It is noteworthy that this luminosity is remarkably close to the one
identified by Fanaroff \& Riley (1974) as the dividing point between
class I and class II type sources. Fanaroff \& Riley also noted
that the luminosity at which the division occurs is remarkably close
to that which divides sources which show strong cosmological evolution
from those which do not (see also Clewley \& Jarvis 2004)

\begin{figure}
\centerline{
\epsfxsize=15cm \epsfbox{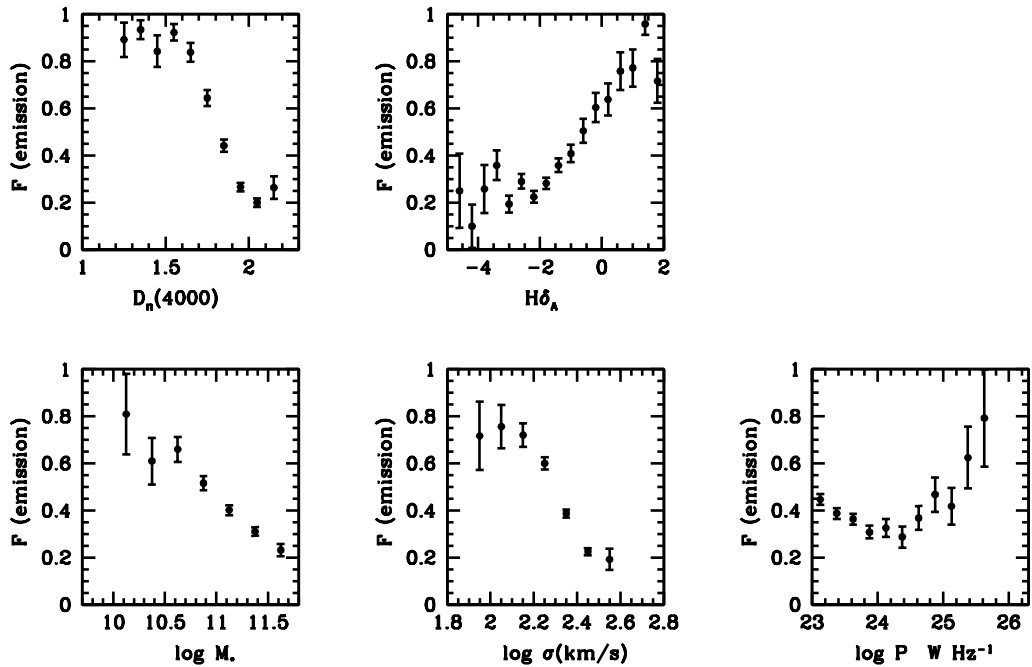}
}
\caption{\label{fig1}
\small
The fraction of radio galaxies with [OIII], H$\beta$, [NII]
and H$\alpha$ detected with $S/N>3$ is plotted as a function
of 4000 \AA\ break,  H$\delta_A$ equivalent width, stellar mass, stellar velocity
dispersion and radio power.}
\end {figure}
\normalsize

\subsection {Radio --  emission line correlations }

The question of whether the emission-line and radio luminosities of
radio AGN are correlated has been examined by a number of groups.
One major problem in these analyses has been the fact that the most
powerful radio galaxies  are always at significantly higher redshifts
than the less powerful radio galaxies in these samples. This makes
it difficult to ascertain whether the primary correlation is between
the emission line luminosity and radio luminosity, rather than between
emission line luminosity and redshift (see McCarthy 1993 for a detailed
discussion). Nevertheless, most studies have favoured the
reality of the correlation (McCarthy 1993; Zirbel \& Baum 1995). 

We now use our sample of low-redshift radio AGN with 
optical emission lines to investigate this issue.
B05b studied  a sample dominated by radio galaxies with weak
or undetected  emission
lines and found no overall correlation. In this section, we
will focus on the subset of radio galaxies with high $S/N$ emission lines. 

As in the previous section, we consider the subsample of radio
AGN with emission lines that are strong enough to allow classification on the
standard BPT diagram ($S/N>3$ in the four emission lines  H$\alpha$, 
H$\beta$, [OIII] and [NII]). The luminosity the [OIII] line
is corrected for extinction using the measured value of the Balmer decrement.
In Figure 4, we investigate the relation between
emission line and radio power for  these objects.
As can be seen, there is a clear 
correlation with substantial scatter. The absence of AGN with very high values of
L[OIII] and low values of $\log P$ may, to some extent,  be a selection effect. We know
that very powerful AGN are usually found in galaxies with strong ongoing
star formation (Kauffmann et al 2003b), so our cut on H$\alpha$ luminosity (see Figure 1) may
eliminate powerful optical AGN with weak radio jets. The absence of galaxies with
low optical line luminosities and very high radio power is, however, a real
effect.  

In the right panel, we restrict
the sample to high ionization optical AGN with $\log [OIII]/H\beta > 0.5$. 
This is the regime occupied by ``pure'' Seyfert galaxies
where there is very little contribution from star formation
to any of the emission lines (Kauffmann et al 2003b). As can be seen,
although the pure Seyferts are boosted to somewhat higher [OIII]
line luminosities than the sample as a whole, they  exhibit much the same    
correlation between [OIII] line luminosity and radio luminosity.
This is not surprising, since it has been demonstrated that there is very little
contribution from star formation to the  [OIII] line luminosity in 
nearby AGN (Kauffmann et al 2003b; Brinchmann et al 2004).

Interestingly, we find that a  substantial part of the scatter in the
relation between [OIII] line luminosity and radio power
 can be attributed to the spread
in the masses and  central velocity dispersions of the 
host galaxies that make up
our sample. In Figure 4, we have colour-coded the points 
according to the value of
the velocity dispersion $\sigma$ measured within the fiber. Cyan points
indicate galaxies with $\sigma$ less than 200 km s$^{-1}$ and red points
indicate galaxies with velocity dispersions larger than this value. For
a given radio power, galaxies with low velocity dispersions
have higher [OIII] line luminosities on average.

We now investigate how the correlation between
radio power and optical line luminosity  for the low redshift SDSS radio galaxy 
sample compares to results  obtained in past studies.
Zirbel \& Baum 1995; hereafter ZB)  carried out the most comprehensive 
study of the relation between radio luminosity and emission line
luminosity in radio galaxies.  They concluded that
FRIs and FRIIs define separate relations with differing slopes. Their data was based
on narrow-band $H\alpha$ imaging, so to be consistent we now plot correlations
using the total measured  luminosity in the 
H$\alpha$ and [NII] lines. In addition, we scale
our results to the same assumed cosmology and radio frequency (408 MHz) used by  
these authors.

In Figure 5,  cyan/red points show EL-RAGN with $\sigma$ values lower
and higher than 200 km s$^{-1}$, respectively. We only plot those galaxies with
$\log$ [OIII]/H$\beta > 0.5$ (as shown in the right-hand panel of Figure 4),
because these ``high-excitation'' radio galaxies  are likely to be more  similar to
the galaxies that ZB classify as FRII sources. 
In addition we have supplemented the sample using radio AGN with emission
lines that were too weak to allow them to be placed on the BPT diagram, but
where H$\alpha$ was still detected with $S/N > 3$ (H$\alpha$ is usually
the line with the highest $S/N$ in the SDSS spectra). These galaxies are
plotted as black points and they should be  good analogues
of the nearby FRI's studied by ZB. The vast majority of  these galaxies 
have high  central velocity dispersions.
For comparison, blue crosses show  radio galaxies from the sample
of ZB that were  classified as FRII or as ``high redshift'' radio sources.
The dashed line shows the  $L(H\alpha)$ -- $\log P$ relation obtained
by ZB for FRII radio galaxies, while the dotted line shows their FRI relation.

It is reassuring that the results for  our weak emission-line radio AGN agree
very nicely with those obtained for the ZB FRI radio galaxies. 
We find that our  weak emission-line AGN span the same range in radio 
luminosities and exhibit the same correlation between
radio luminosity and emission-line luminosity found by these authors.  
Our stronger emission-line radio AGN appear to  lie on the 
low-luminosity  extrapolation of the ZB relation for FRII galaxies.
It is also worth noting that radio galaxies
with high values of $\sigma$ appear to lie close to a   {\em direct} extrapolation of
this relation, whereas local radio galaxies with low values of $\sigma$  
are offset to lower values of $\log P$.
The results suggest that the more powerful FRII radio
galaxies may arise when galaxies with high mass black holes 
accrete at a significant rate. Such systems are very rare in the
low redshift Universe, but become increasingly more common at earlier
epochs.

\begin{figure}
\centerline{
\epsfxsize=14cm \epsfbox{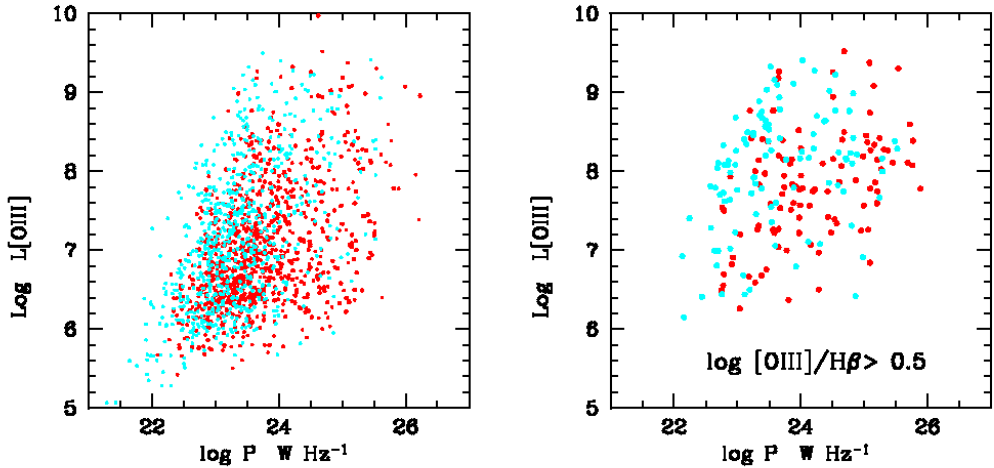}
}
\caption{\label{fig1}
\small
Left: Correlations between emission line and radio strength for the EL-RAGN sample. Cyan
points indicate radio AGN  with $\sigma < 200$ km/s and red points indicate  radio
AGN with $\sigma > 200$ km/s. Right: Results are shown for the subset of
EL-RAGN with high ionization emission lines. Note that L[OIII] is in solar units.
The radio luminosities are defined at 1.4 Ghz. }

\end {figure}
\normalsize

\begin{figure}
\centerline{
\epsfxsize=13cm \epsfbox{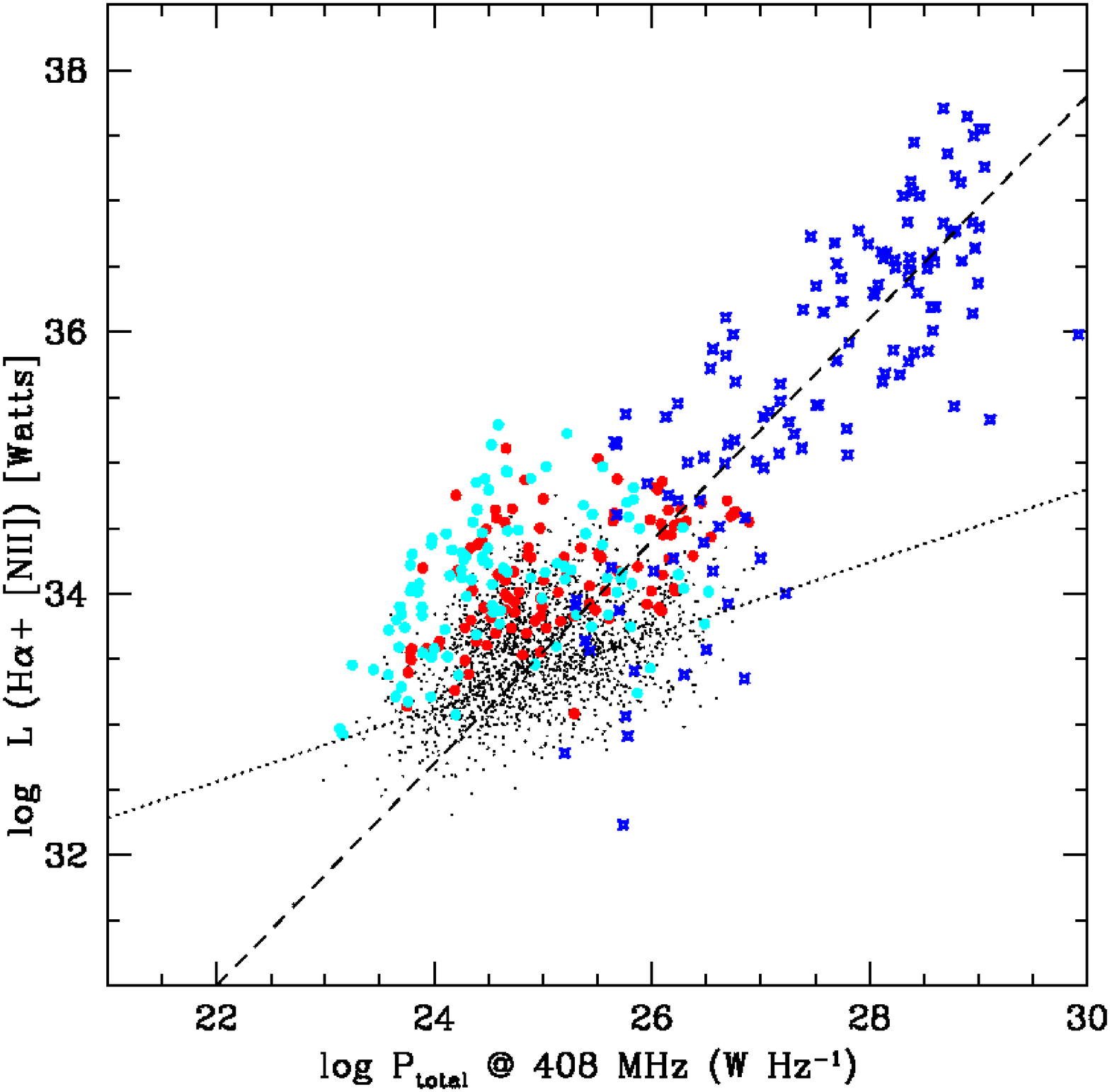}
}
\caption{\label{fig1}
\small
Comparison between our results and those of Zirbel \& Baum. The sum of the H$\alpha$ and [NII]
emission line luminosities is plotted as a function of radio luminosity. Results have been
scaled to a cosmology with $H_0 =50$ km s $^{-1}$ Mpc$^{-1}$ and to radio luminosities measured
at 408 MHz (see text). Cyan and red points are the sample of low-$\sigma$ and high-$\sigma$
EL-RAGN plotted in the right-hand panel of Figure 4 (i.e. they are ``high-excitation''
sources with $\log [OIII]/H\beta > 0.5$) . Black points are radio AGN where H$\alpha$+
 [NII] is detected with $S/N >3$, but other emission lines are too weak to allow
the galaxy to be classified on the BPT diagram.   FRII and high-redshift
radio galaxies from the sample of Zirbel \& Baum (1995) are plotted as blue crosses. 
The dashed and dotted lines indicate
the Zirbel \& Baum  fits to  the FRII and FRI radio galaxies in their sample.}

\end {figure}
\normalsize

We have shown that for nearby radio AGN, the relation between
emission line luminosity and radio luminosity depends on the velocity
dispersion of the galaxy. 
In order to gain some  physical understanding of this trend, 
we scale both the radio and the [OIII] luminosities of the AGN in
our sample by dividing by the 
``black hole mass'', which can be estimated from $\sigma$ using the
relation given in Tremaine et al (2002).
As discussed in Heckman et al (2004), L[OIII]/M$_{BH}$ may be 
regarded as a measure of the accretion rate onto the black hole in
Eddington units, and measures how rapidly black holes have been 
building up their present-day mass. Heckman et al used
this indicator to demonstrate that galaxies with low values
of $\sigma$ (and hence low mass black holes) are currently building up their
black holes at a much higher rate in comparison 
to galaxies with high values of $\sigma$.  

In the right hand panel of  Figure 6, we plot the  relation between the scaled [OIII] and radio
luminosities. As before, emission-line
AGN with $\sigma < 200$ km s$^{-1}$ are plotted as cyan points and 
AGN with $\sigma > 200$ km s$^{-1}$ are plotted in red.
We again find a clear correlation between the scaled quantities          
 with substantial scatter.
In agreement with
results of Heckman et al (2004), there is an offset in scaled
[OIII] luminosity  between AGN with high/low values of $\sigma$.
Interestingly, however, the offset in radio luminosity disappears
after it is  scaled by the black hole mass.
This  result is shown more quantitatively
in Figure 7, where we plot  the median, lower 25th percentile
and upper 75th percentile of the  radio luminosity at a given value of  L[OIII]
(left panel) and of  the  scaled radio luminosity at a given value of  of L[OIII]/$M_{\rm BH}$ (right panel).
In contrast to Figure 6,  results are shown only for the RL-AGN sample.
Cyan and red lines indicate galaxies with $\sigma < 200$ km s$^{-1}$ 
and $\sigma > 200$ km s$^{-1}$, respectively.   
The correlation between
normalized radio power and the accretion rate in Eddington units  is nearly
independent of black hole mass. However, the high
mass (low mass) black holes dominate the distribution at low (high)
normalized luminosities and accretion rates.

For completeness and to compare with the results of B05b, we also plot
radio AGN without high S/N emission lines as black points in Figure 6.
The L[OIII] values correspond to upper limits in the cases where
the [OIII] line was not detected, or was detected with marginal significance.
As can be seen,  radio AGN with weak emission lines are most numerous               
at all but the very highest radio luminosities. The yellow
dashed line indicates the median value of L[OIII]
as a function of
radio luminosity  for the whole sample.
There is essentially no trend in the median  [OIII]
luminosity of the radio AGN  up to  a radio luminosity
of $10^{25}$ W Hz$^{-1}$, where  there is a slight 
upturn (see also B05b). This upturn is considerably 
more pronounced if the scaled radio luminosity is used.    
Radio AGN with weak emission lines are  absent  above
a scaled radio luminosity of $10^{17}$ W Hz$^{-1}$ M$_{\odot}^{-1}$.
In the local Universe, most such objects have low black hole masses
and thus do not have extremely high total radio luminosities.
At high redshifts, however, we speculate that high mass black holes will be actively
accreting and typical total radio luminosities will thus be orders of magnitude
higher than in our SDSS sample.

Finally, we study whether there is a correlation between radio power
and the stellar populations of the radio AGN in our sample.
 Figure 8 shows that there is no correlation between H$\delta_A$
and radio luminosity, but there is a  trend towards younger mean stellar ages
at high values of the scaled radio luminosity $\log P$/ M$_{BH}$.
The majority of the radio AGN with young stellar population have low
velocity dispersions. Once again we see that the {\em median} value of
H$\delta_A$  is set by the large number of old systems 
in our sample and remains constant  up to  a scaled radio luminosity of 
$\sim 10^{17}$ W Hz$^{-1}$ $M_{\odot}^{-1}$, where it exhibits an abrupt
upturn.

\begin{figure}
\centerline{
\epsfxsize=14cm \epsfbox{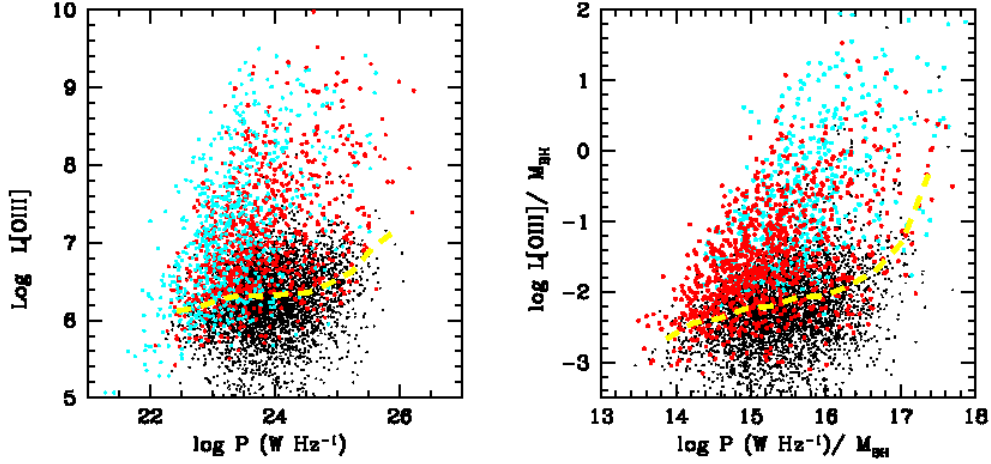}
}
\caption{\label{fig1}
\small
Correlations between emission line and radio luminosities  for the full  sample. Cyan
points indicate radio AGN  with emission lines and with $\sigma < 200$ km/s; red points indicate  radio
AGN with emission lines and with $\sigma >$ 200 km/s;  black points indicate radio AGN with weak or absent
emission lines (in the case of non-detections, upper limits are
plotted). In the left panel [OIII] luminosity is plotted as a function of radio
luminosity. In the right panel, both the [OIII] and the radio luminosities are scaled by dividing by the
black hole mass. The yellow dashed line shows the median value of L[OIII] or L[OIII]/$M_{BH}$
as a function of  radio luminosity.}

\end {figure}
\normalsize

\begin{figure}
\centerline{
\epsfxsize=12cm \epsfbox{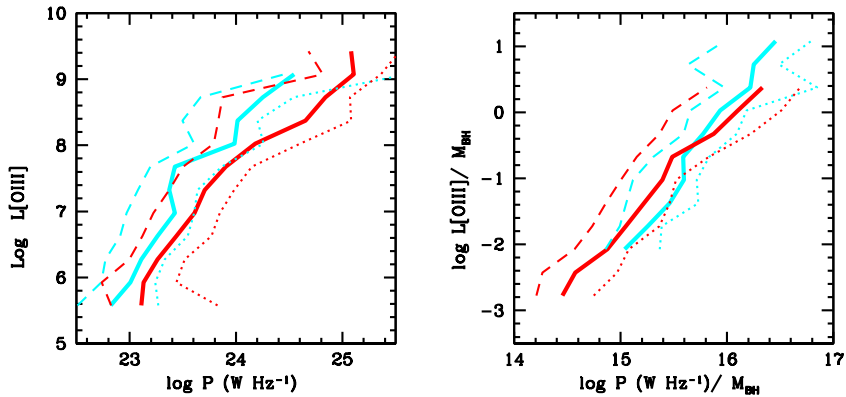}
}
\caption{\label{fig1}
\small
The figure shows the  median (solid), lower 25th percentile (dashed) and
upper 75th percentile (dotted) of the  radio luminosity at a given value of  L[OIII]
(left),  and the same for the  scaled radio luminosity at a given value of  L[OIII]/$M_{\rm BH}$ (right).
In contrast to Figure 6, results are shown only for the EL-RAGN sample.
The cyan curves are for galaxies with $\sigma < 200$ km s$^{-1}$ while the red curves
are for galaxies with $\sigma > 200$ km s $^{-1}$.}
\end {figure}
\normalsize

\begin{figure}
\centerline{
\epsfxsize=14cm \epsfbox{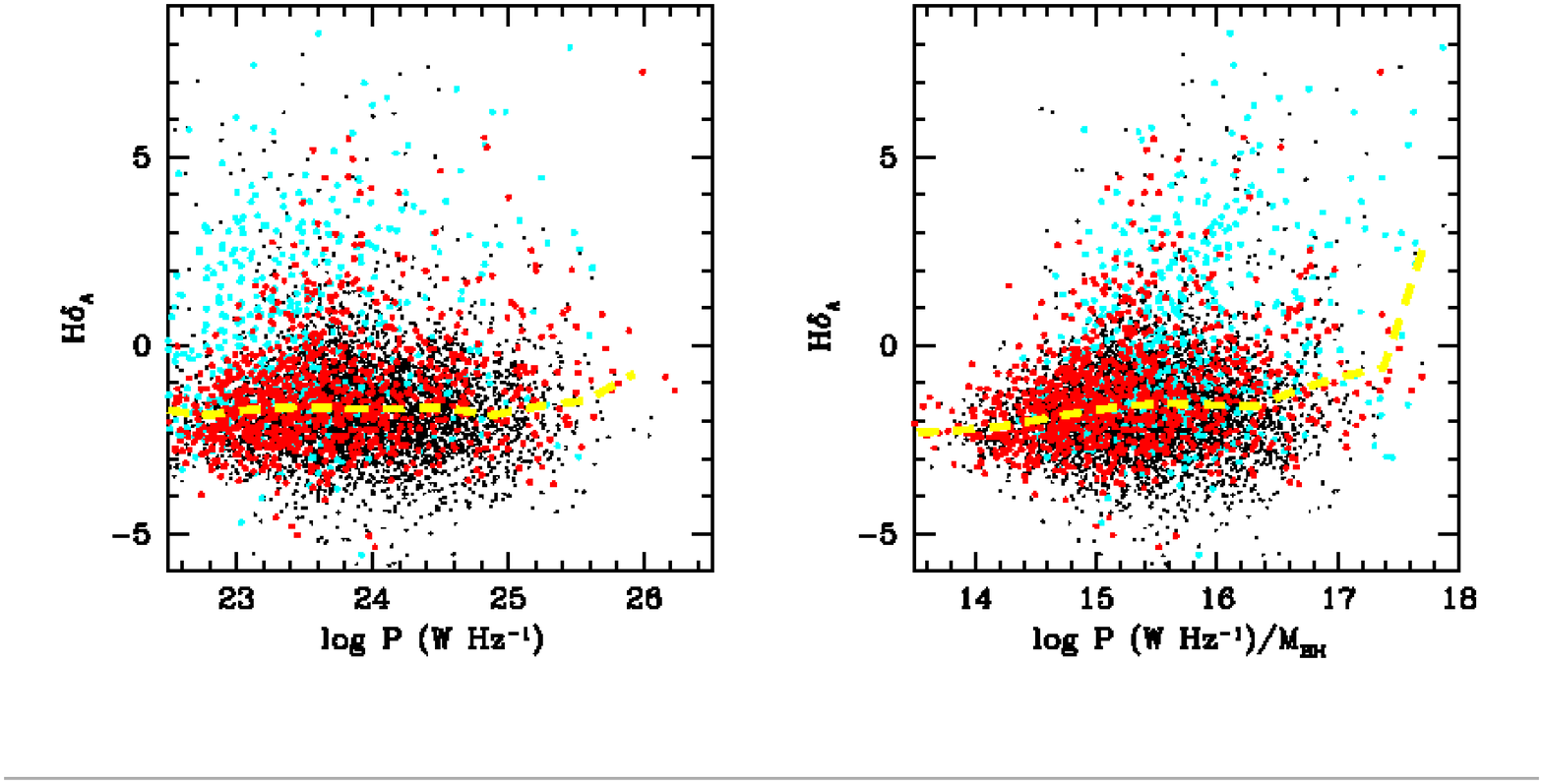}
}
\caption{\label{fig1}
\small
Correlations between H$\delta_A$ equivalent width  and radio luminosity  for the full sample. The colour
coding of the points is the same as in Figure 6.}

\end {figure}
\normalsize

\subsection {Section summary}

Radio-loud AGN are most likely to display emission lines in galaxies with low stellar
masses and velocity dispersions, and at radio luminosities greater than 
$10^{25}$ W Hz$^{-1}$. The radio and emission line luminosities are correlated, but with 
substantial scatter. At a fixed radio luminosity, AGN in low velocity dispersion
hosts tend to  have higher [OIII] luminosities than AGN in high velocity dispersion hosts.
However, if we scale the radio and [OIII] luminosities by dividing by the black
hole mass estimated from  our $\sigma$ measurements, we find a coorelation between
normalized radio luminosity and  
L[OIII]/$M_{BH}$ that is independent of black hole
mass. Scaling the radio luminosity by the black hole mass also 
uncovers a correlation between the normalized 
radio power and the age of the stellar population
in the host galaxy.

\section { How do radio-loud AGN with emission lines 
differ from  radio-quiet AGN?}

\subsection {Host galaxies}
Unlike the low excitation radio sources, which are known to occur
in normal massive early-type galaxies with very little ongoing star formation,
the more powerful radio galaxies with strong emission lines
have been found to occupy more peculiar hosts. Smith \& Heckman (1989) found that
half of the powerful radio galaxies in their sample exhibited peculiar
optical morphologies (tails, fans, bridges, shells and dust lanes).
Their surface brightness profiles were also shallower than those of normal
elliptical galaxies and many of the objects also had bluer than average
colours compared to giant ellipticals. More recently, Tadhunter et al (2005)
analyzed high-quality long-slit spectra  for a sample of nearby
powerful radio galaxies and showed that all of them had a significant
young stellar component. 

Care must be exercised, however, in interpreting these results as implying
that the radio jet is {\em caused} by the same process that causes
the enhanced star formation or the  morphological peculiarities.
Optical AGN activity has also been shown to be associated with
enhanced levels of star formation in a galaxy (Kauffmann et al 2003b;
Wild et al 2007). 
In order to disentangle the physical processes that control           
the radio emission from those that control the optical
emission, we introduce a pair-matching technique. For each EL-RAGN, we find
a radio-quiet emission line AGN (henceforth RQAGN) 
that is  closely matched in redshift, in stellar mass and in 
velocity dispersion. The motivation for choosing this set of 
matching parameters is the following: 
\begin {enumerate}
\item By constraining the matching galaxy to lie
at the same redshift, we make sure that the fiber apertures subtend
the same physical scale and that the  both the radio luminosity
and the optical emission line luminosity are subject to the same selection biases
caused by the survey flux limits.   
\item By constraining the matching galaxy to have the same  
velocity dispersion, we make sure that both galaxies
have roughly the same black hole mass.  Any AGN  
physics that is controlled  primarily by the black hole mass will then be the same.
\item By constraining the matching galaxy to have the same stellar mass, we make sure
that the host galaxies of both types of AGN are drawn from 
roughly similar parent galaxy populations. 
\end {enumerate}
Because the underlying sample of radio quiet AGN is extremely large, 
we have no problem imposing matching tolerances as
stringent as $\Delta z < 0.01$, $\Delta \log \sigma < 0.05$   
and $\Delta \log M_* < 0.1$.

Figure 9 presents a comparison of the host galaxy properties of our EL-RAGN sample
(red histograms) and the matched sample of RQAGN (black histograms).
The galaxy properties that we compare include
the distributions of  concentration, stellar
surface mass density $\mu_*$, 4000 \AA\ break strength,
H$\delta$ absorption line equivalent width, H$\alpha$ emission line
equivalent width, and the ``Eddington'' parameter L[OIII]/$M_{BH}$.  Note that
there are around 1600 galaxies in both these samples, so the shot-noise error is small.
As can be seen,  the radio loud and  matched radio quiet galaxies have nearly
identical stellar populations and emission line properties. The  property
that differs the most between the two samples 
is the concentration index: the host galaxies of  EL-RAGN appear to
be slightly  more concentrated than those of their radio-quiet counterparts. 

\begin{figure}
\centerline{
\epsfxsize=14cm \epsfbox{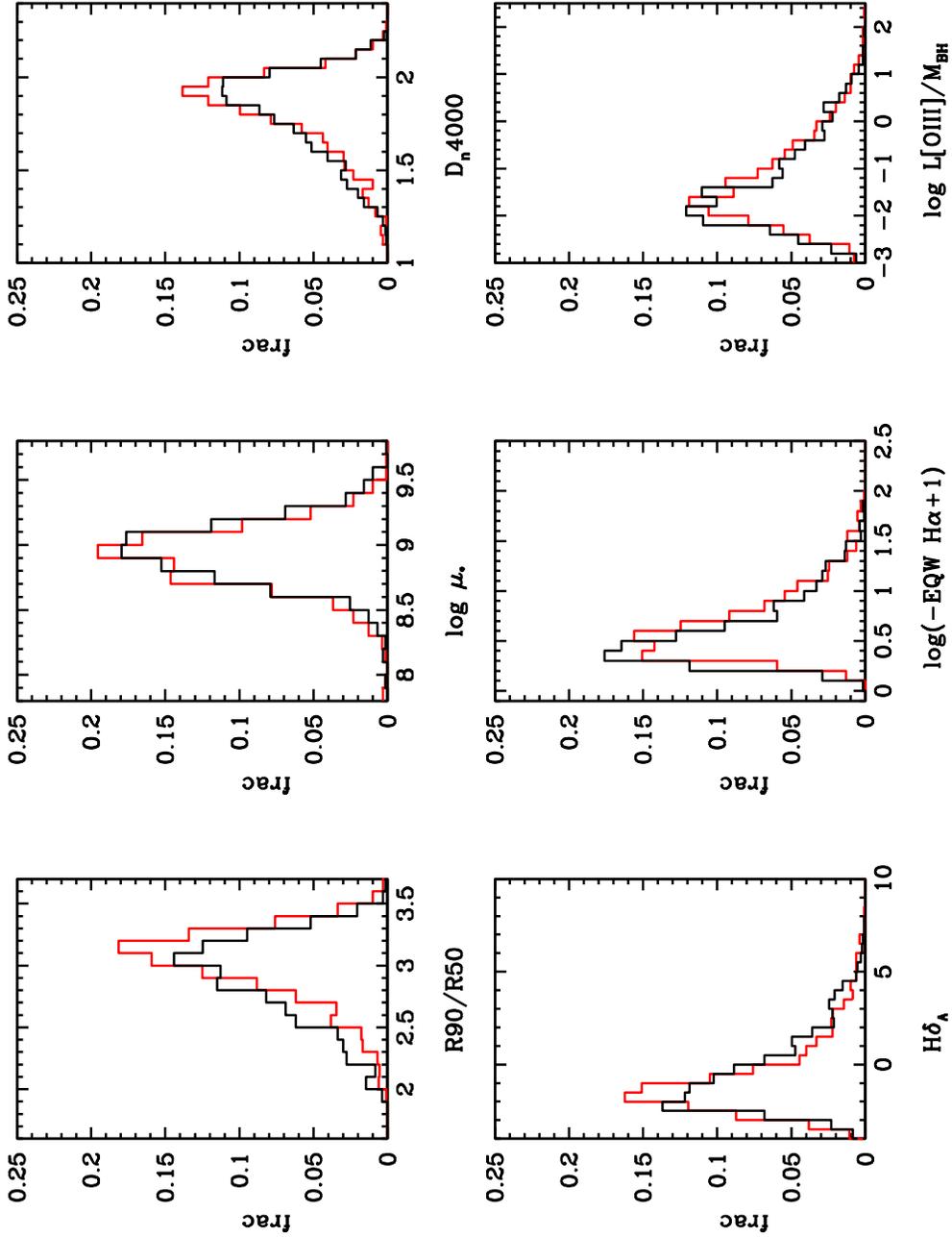}
}
\caption{\label{fig1}
\small
Comparison of host galaxy properties of EL-RAGN (red) and matched RQAGN (black).
We plot histograms of concentration index (R90/R50), stellar surface mass density
in units of $M_{\odot}$ kpc$^{-2}$, 4000 \AA\ break strength, H$\delta_A$ index,
H$\alpha$ equivalent width, and  [OIII] luminosity normalized by black
hole mass.}

\end {figure}
\normalsize

\subsection {Environment}
We have also investigated the environments of radio-loud 
and radio-quiet emission line AGN.
We characterize environment by computing galaxy  counts down to a limiting $i$-band
magnitude of 20.0 around each galaxy, using the SDSS photometric catalogue. 
The counts are evaluated as a function of angular separation out to a limiting
distance of 3 arcmin. By binning together many different galaxies, we reduce
the noise in this estimator. We also make a statistical correction for galaxies
that are not physically associated with the AGN by computing counts around 
randomly distributed sky coordinates and subtracting the average uncorrelated count
from our local density estimate.

Figure 10 presents results for  the EL-RAGN sample. In the left panel,
we plot the average number of galaxies interior to radius $R$ . The 
cyan, green, red and
magenta lines are for EL-RAGN binned according to increasing radio luminosity.
As can be seen, the environments do not change significantly as a function of
radio luminosity. This should be contrasted with the right-hand panel, in which we
have binned the EL-RAGN according to velocity dispersion $\sigma$. It is clear
that radio AGN with bigger bulges and black holes reside 
in systematically denser environments.
We note that there is no trend in the
median redshift of our sample of EL-RAGN as a function of radio luminosity
and only  a small increase  in $\bar{z}$  from 0.096  for
galaxies in the lowest $\sigma$ bin compared to 0.132 for  galaxies in the highest $\sigma$ bin.
Our conclusion that clustering does not depend on radio power is thus quite robust. 
If we were to correct for the small increase in redshift as a function of
$\sigma$, this would slightly boost the environmental differences that are
already quite clear from our plot.

In Figure 11, we compare the environments of the EL-RAGN with the matched sample
of RQAGN. We show results in four different ranges of central velocity
dispersion of $\sigma$. We find that for all values of $\sigma$, the galaxy counts
around the radio-loud objects are a factor of $\sim 2$ higher than around their radio
quiet counterparts. This is in remarkable agreement  
with the results of Smith \& Heckman (1990) who compared the 100 kpc
scale environments of a sample of 31 low redshift ($z< 0.3$)  QSOs with
a sample of 35 powerful radio galaxies with $\log P > 10^{25}$ W Hz$^{-1}$.
These authors found that both the radio-loud QSOs and radio galaxies had 
approximately twice the average number of close companions  
as compared to the radio-quiet QSOs.

\begin{figure}
\centerline{
\epsfxsize=12cm \epsfbox{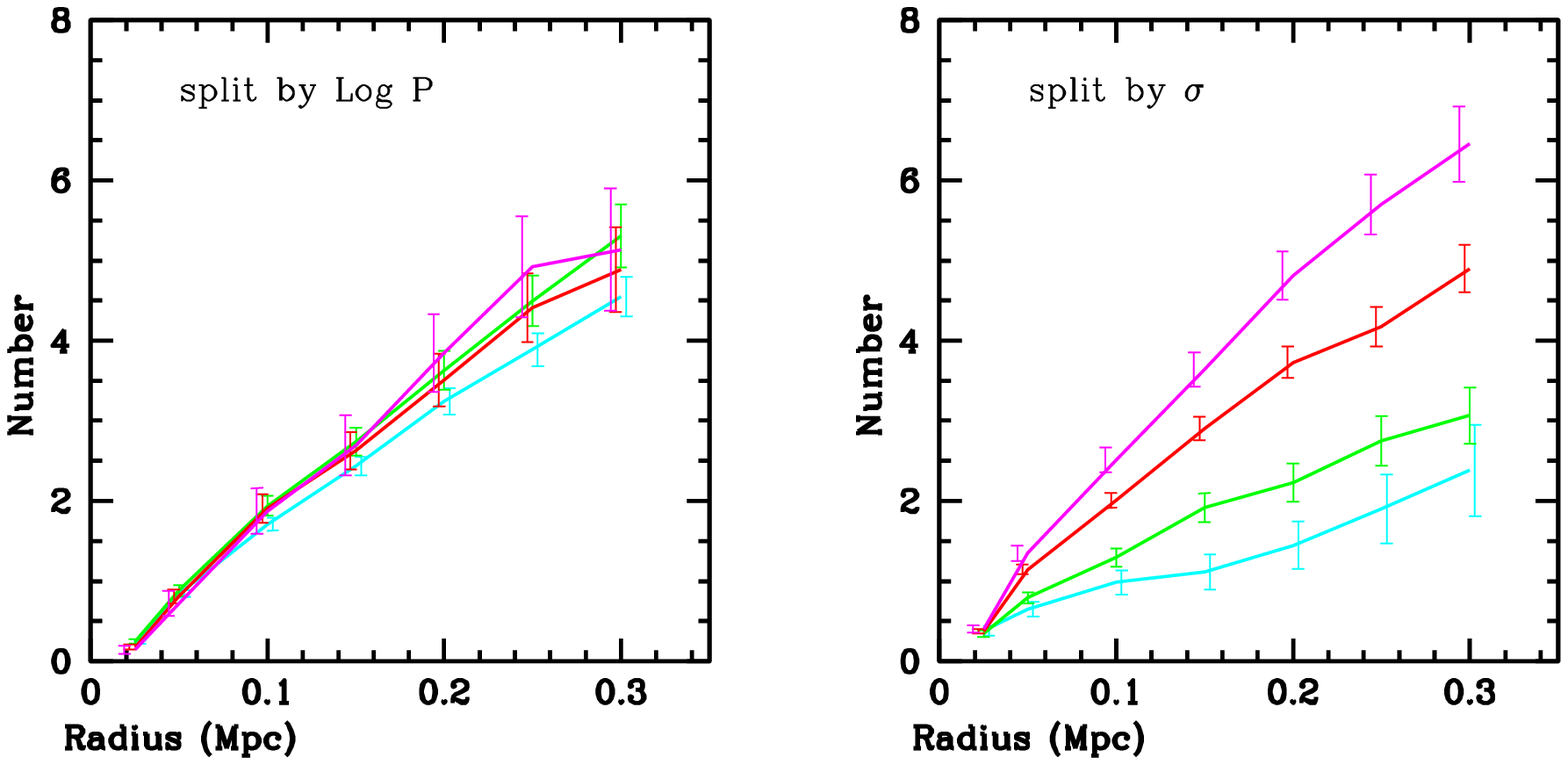}
}
\caption{\label{fig1}
\small
Galaxy counts within  projected radius $R$ in Mpc  for EL-RAGN. In the left
panel,  the sample
is split according to radio luminosity and in the right panel, the sample is split according to 
velocity dispersion $\sigma$. In both panels, 
blue, green, red and magenta lines show results for
subsamples ordered in increasing radio power (left) or velocity dispersion (right). }
\end {figure}
\normalsize

\begin{figure}
\centerline{
\epsfxsize=10cm \epsfbox{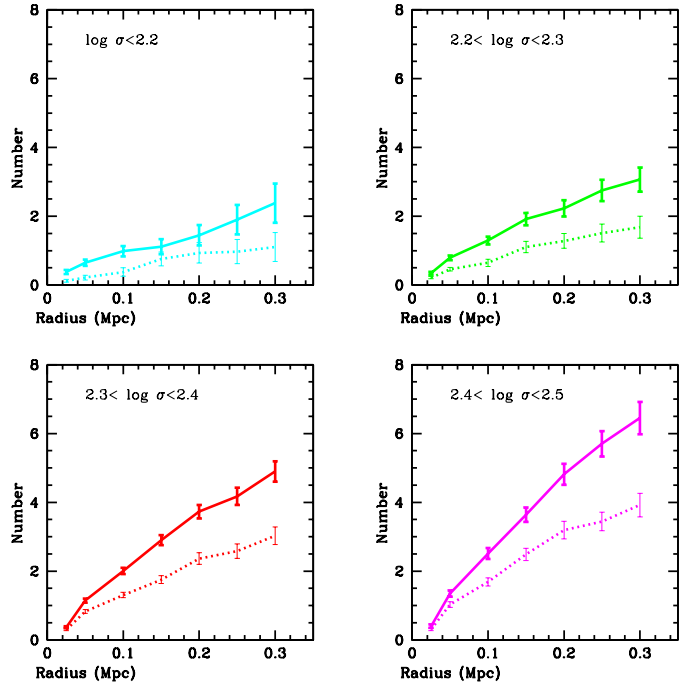}
}
\caption{\label{fig1}
\small
Galaxy counts within projected  radius $R$ in Mpc  for EL-RAGN (solid)
and matched samples of radio-quiet AGN (dotted). Results are shown
for 4 different ranges
in velocity dispersion  $\sigma$.}
\end {figure}
\normalsize

\subsection {Emission line properties}
In this section, we investigate whether there are any detectable 
differences in the ionization state of the narrow-line regions of
the galaxies in EL-RAGN and the matched radio-quiet samples. In order to
do this, we impose one further matching constraint on the radio quiet sample:
we require that the value of extinction corrected [OIII] 
luminosities match to better than 0.15 dex.

Our results are illustrated in Figure 12. In the left-hand panel, 
we plot the 4000 \AA\ break index as 
a function of the normalized [OIII] line luminosity L[OIII]/M$_{BH}$. This probes
the correlation between star formation and accretion rate onto the black hole.
EL-RAGN  are plotted in red and the radio-quiet  matches are plotted in blue.
In the middle and right-hand panels, we compare the positions of both kinds
of galaxies on two of the standard line diagnostic (BPT) diagrams. Results are
shown in three different ranges of radio luminosity. 

Figure 12 shows that to first order, EL-RAGN and their 
radio quiet counterparts are remarkably
similar in terms of their emission line properties. 
There are some small differences,
which are quantified in detail in Figure 13. In the left panel, we plot
the distribution of D$_n$(4000) values for the EL-RAGN with radio luminosities in
excess of $3 \times 10^{23}$ W Hz$^{-1}$ in red, and the distribution of
D$_n$(4000) for their radio-quiet counterparts in blue. As can be seen, the
EL-RAGN are shifted to somewhat higher values of 
D$_n$(4000) Recall that the red and blue
samples are exactly matched in  L[OIII], so
this result implies that radio-loud objects have slightly older 
stellar populations at a fixed
value of L[OIII].  

In Kauffmann et al (2003b), we introduced two parameters for 
characterizing the position
of an AGN in the BPT diagram: a) a distance parameter $D$, 
which measures how far away
the galaxy is from the star-forming locus, 
b) an angle parameter $\Phi$, which characterizes
the ionization state of the galaxy (i.e. the boost in [OIII]/H$\beta$ relative to
[NII]/H$\alpha$.) The middle and right panels of Figure 13 compare the
distributions of $D$ and $\Phi$ for our two samples. We find that EL-RAGN have
somewhat larger values of $D$, but essentially identical values of $\Phi$
when compared to the radio quiet control galaxies. This is again consistent with
the notion that the radio-loud AGN have slightly older stellar populations on
average.

\begin{figure}
\centerline{
\epsfxsize=14cm \epsfbox{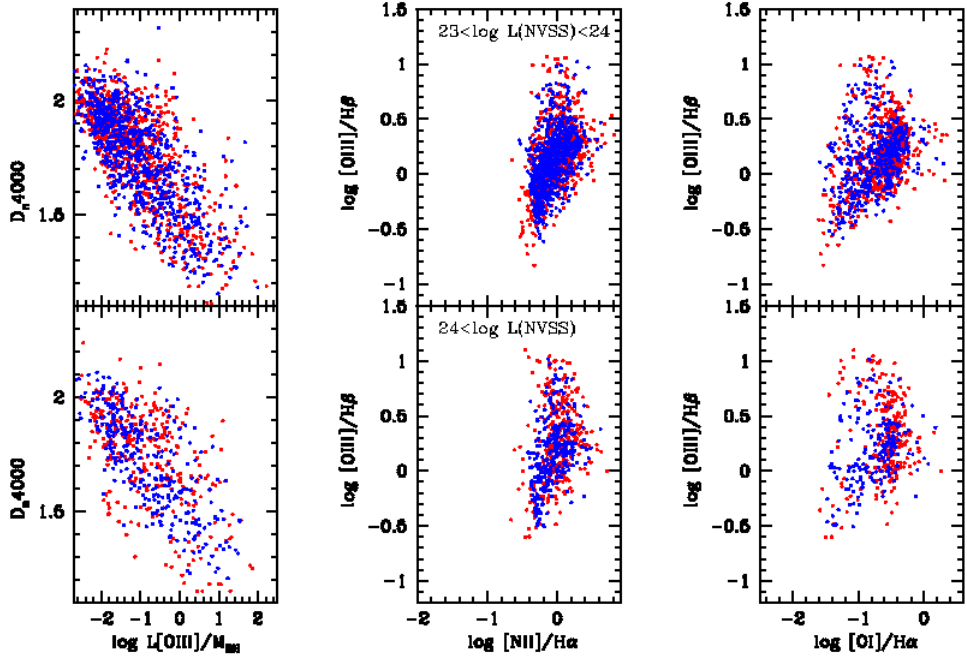}
}
\caption{\label{fig1}
\small
Comparison of emission line properties for EL-RAGN (red) and their radio quiet counterparts (blue).
The left panels show the distribution of 4000 \AA\ break strength as a function
of normalized [OIII] line luminosity. The two right panels show two different
examples of BPT (Baldwin, Phillips \& Terlevich 1981) diagrams.} 
\end {figure}
\normalsize

\begin{figure}
\centerline{
\epsfxsize=16cm \epsfbox{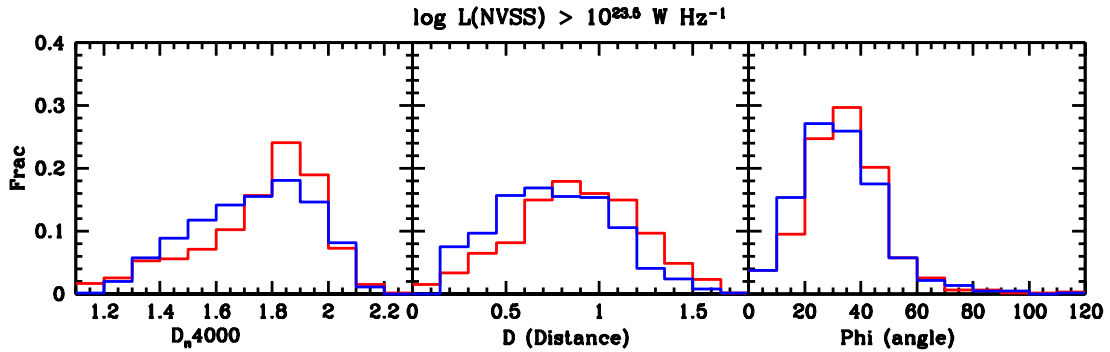}
}
\caption{\label{fig1}
\small
Quantitative comparison of emission line properties for EL-RAGN (red histogram) and their 
radio quiet counterparts (blue histogram). We plot distributions of 4000 \AA\ break strength,
as well as the parameters $D$ and $\Phi$, which quantify the location of the
galaxy in the [OIII]/H$\beta$ versus [NII]/H$\alpha$ BPT diagram (see text).}
\end {figure}
\normalsize

\subsection {Section Summary}

Overall, we find  a remarkable  similarity between the radio-loud
and the matched radio-quiet objects in terms of host galaxy structure, 
stellar populations and emission line properties. Radio-loud AGN with 
emission lines appear to be  more bulge-dominated and to have
older stellar populations than their radio quiet counterparts,
but the differences are very small. 
The most striking difference between the two types of AGN is in 
the environments in which they are located. We find a factor of two difference
in local density between the two populations. 
Interestingly, this difference in environment only shows up
when radio-loud AGN are compared with radio-quiet AGN.
There is no correlation between the {\it radio  luminosities}
of the galaxies in the EL-RAGN sample and their environment. 
Best et al (2007) found a similar result for radio AGN in groups
and clusters. The probability for a galaxy to be radio loud was
higher in the central group/cluster galaxy when compared to other galaxies of the
same stellar mass, but the distribution of radio luminosities
did not depend on location within the cluster.
This is somewhat puzzling, because simple jet models predict that radio
galaxies should be more luminous in denser media,
because adiabatic expansion losses are reduced and more energy is radiated
(e.g. Kaiser, Dennett-Thorpe \& Alexander 1997).

The result that radio-loud AGN with emission lines reside
in denser environments than radio-quiet emission line AGN 
is in good agreement with 
the  conclusions presented by Smith \& Heckman (1990). It also agrees with
conclusions about differences in the  environments of 
radio-loud and radio-quiet quasars
by Yee \& Green (1984) and Ellingson et al (1991), but does not
agree with later work by McLure \& Dunlop (2001) on this topic.

\section {Which processes trigger the optical and radio AGN phenomena?}

The results presented in the previous section suggest that a high density
environment is required in order to produce a  radio jet that
is sufficiently luminous to be included in our catalogue. On the
other hand, analyses
of optically-selected AGN  suggest that AGN with strong emission lines
are almost always associated with ongoing star formation in the galaxy
and, by extension, a reservoir of cold gas. Putting these two
results together, {\em we postulate that radio-loud AGN with emission
lines occur in the small subset of galaxies that contain sufficient cold
gas to power strong emission lines, but also reside in high density environments.}

In the local Universe, low mass galaxies are generally gas-rich 
and actively star-forming, while the highest mass galaxies are almost always
gas-poor and have very little ongoing star formation. 
This is probably  why optical AGN  with the highest values of $L/L_{Eddington}$  
tend to occur in galaxies with the smallest bulges and black
holes (Heckman et al 2004). The majority of such galaxies are 
found in low density environments,
which are not conducive to the formation of radio-emitting jets.
The requirement of high local density pushes the EL-RAGN into the
peculiar parameter space of galaxies with relatively low masses that
are in denser-than-average environments. 

These arguments can be made more concrete by considering  radio and strong 
emission line AGN separately  and asking how they 
compare to matched samples of galaxies
that are selected without regard to their AGN properties.
Once again we choose to match in redshift, stellar mass and 
velocity dispersion. The radio AGN sample consists of     
all 5714 sources in the DR4 area with NVSS fluxes greater than 2.5 mJy
that are classified as an AGN rather than star-forming (see section 2).
The optical AGN sample includes 4680 objects
that are selected to have $\log$ L[OIII]/M$_{BH} > 1$. This
cut selects approximately 
5\% of all the AGN in our sample with the most rapidly growing
black holes. According to
the calibration between [OIII] line luminosity and accretion rate
given in Heckman et al (2004), and taking into account the average extinction
correction that we apply to the [OIII] line in this analysis, 
the black holes in these galaxies are    
accreting at approximately a tenth of  Eddington or more. 

Figure 14 compares the distributions of 4000 \AA\ break 
strength and concentration index
for radio AGN (red histograms) with matched samples of  galaxies
selected without regard to AGN properties (black).
Results are shown for  four different ranges of central 
velocity dispersion $\sigma$.

We see that the largest differences arise  for the low velocity dispersion systems.
Radio-loud AGN with low values of $\sigma$  are shifted to higher values 
of both D$_n$(4000) and concentration index compared to their matched
galaxies. At the largest velocity dispersions, the distributions
are almost identical. This is not surprising, because the fraction of 
galaxies that are radio loud approached values close to unity at the
very highest masses and velocity dispersions (Best et al 2005).

Figure 15 compares the distributions of 4000 \AA\ break strength 
and concentration index for our sample of powerful 
optical  AGN (blue histograms) with matched samples galaxies
selected without regard to AGN properties (black).
It is clear  that the trends are quite different to what 
is seen for the radio AGN.
Strong differences arise at all velocity dispersions, but only in the
properties of the stellar populations of the optical AGN,
not in their structural properties.
We note that these results are fully in accord with 
the work of  Kauffmann et al (2003b),
who concluded that powerful optical AGN reside in galaxies that are structurally
similar to ordinary early-type systems, but have stellar populations that
are considerably younger.

\begin{figure}
\centerline{
\epsfxsize=15cm \epsfbox{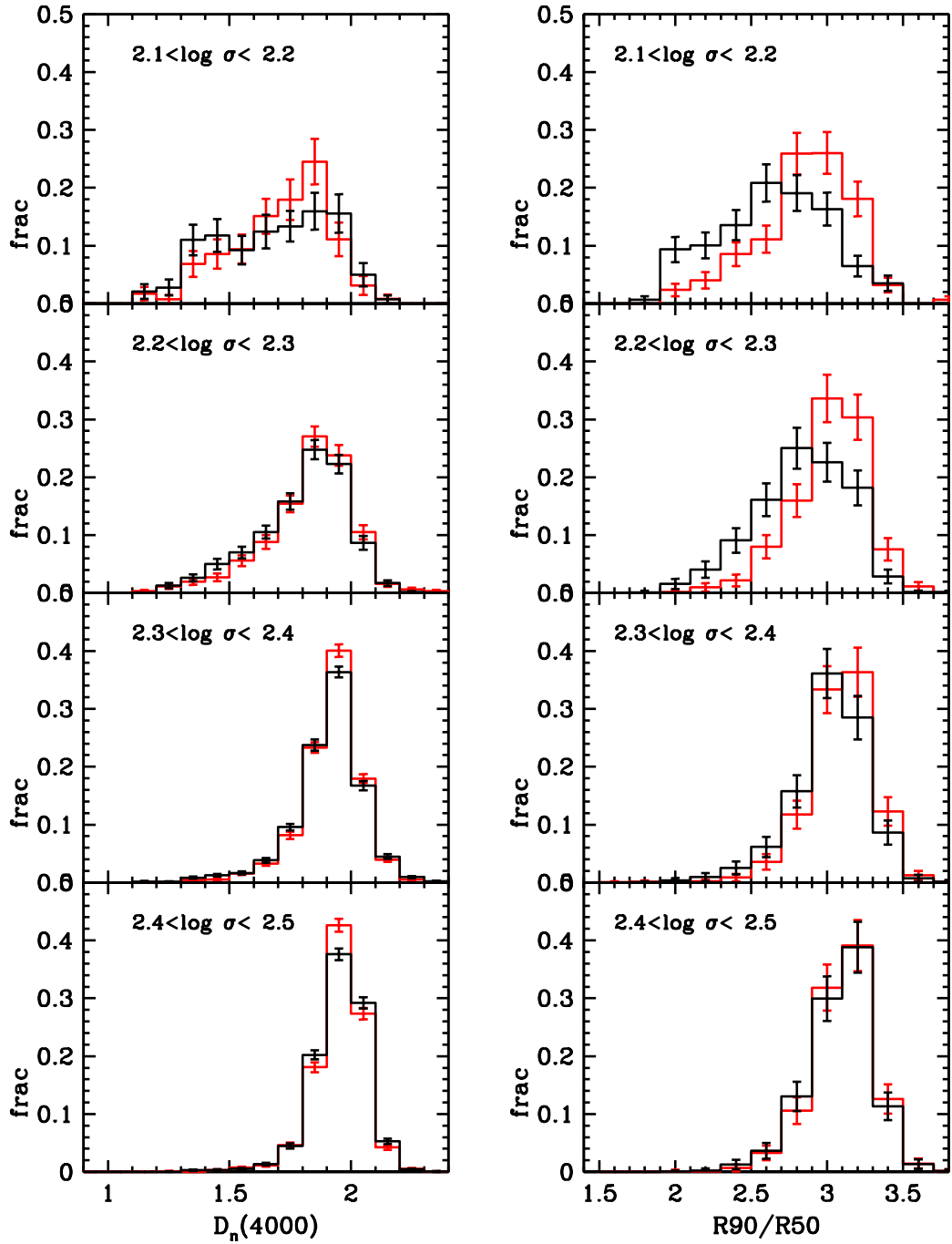}
}
\caption{\label{fig1}
\small
Comparison of the D$_n$(4000) and concentration index distributions
of our sample of radio-loud AGN (red) and a matched  sample of galaxies selected without regard
to AGN properties (black).}   

\end {figure}
\normalsize

\begin{figure}
\centerline{
\epsfxsize=15cm \epsfbox{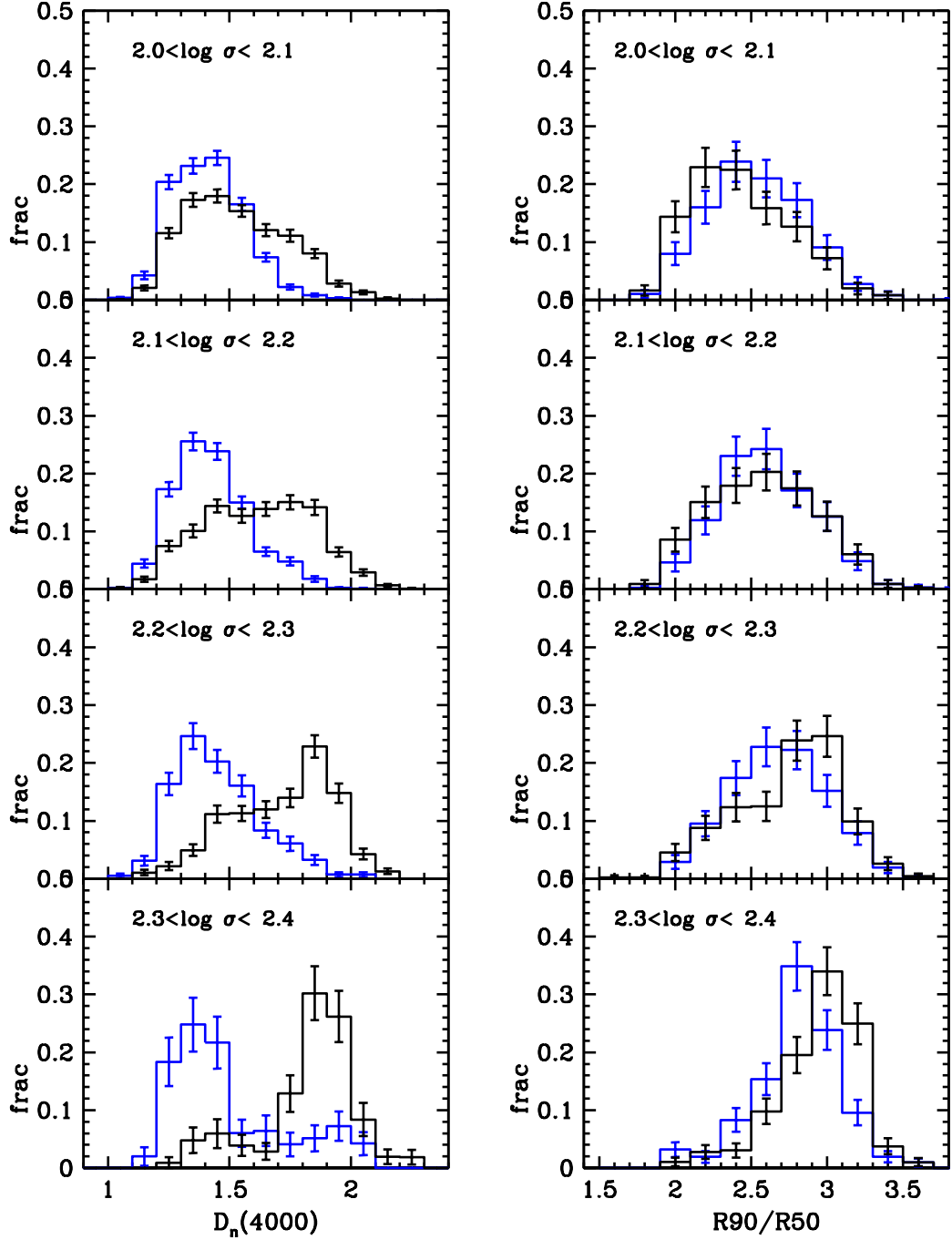}
}
\caption{\label{fig1}
\small
Comparison of the D$_n$(4000) and concentration index distributions
of our sample of our sample of powerful optical  AGN with high
L[OIII]/M$_{BH}$  (blue) and a matched  sample of galaxies selected without regard
to AGN properties (black).}   
\end {figure}
\normalsize

We now compare the environments of the two classes of AGN with matched  control
samples of  galaxies that are selected without regard to their
AGN properties. The top panel of Figure 16 shows that radio AGN
are located in denser environments than the control galaxies. 
Once again the differences are substantially 
larger for the lower velocity dispersion systems. In contrast, 
powerful optical AGN are always located in similar or slighty 
less dense environments that the control galaxies. The result on
optical AGN is in agreement with the analysis of the clustering of 
optical AGN presented in Li et al (2006). The trends found for radio AGN agree  
with the results presented in   Best et al (2007), who analyzed           
central group and cluster galaxies and found that they have a higher probability of being
radio loud than other galaxies of the same stellar mass.

\begin{figure}
\centerline{
\epsfxsize=15cm \epsfbox{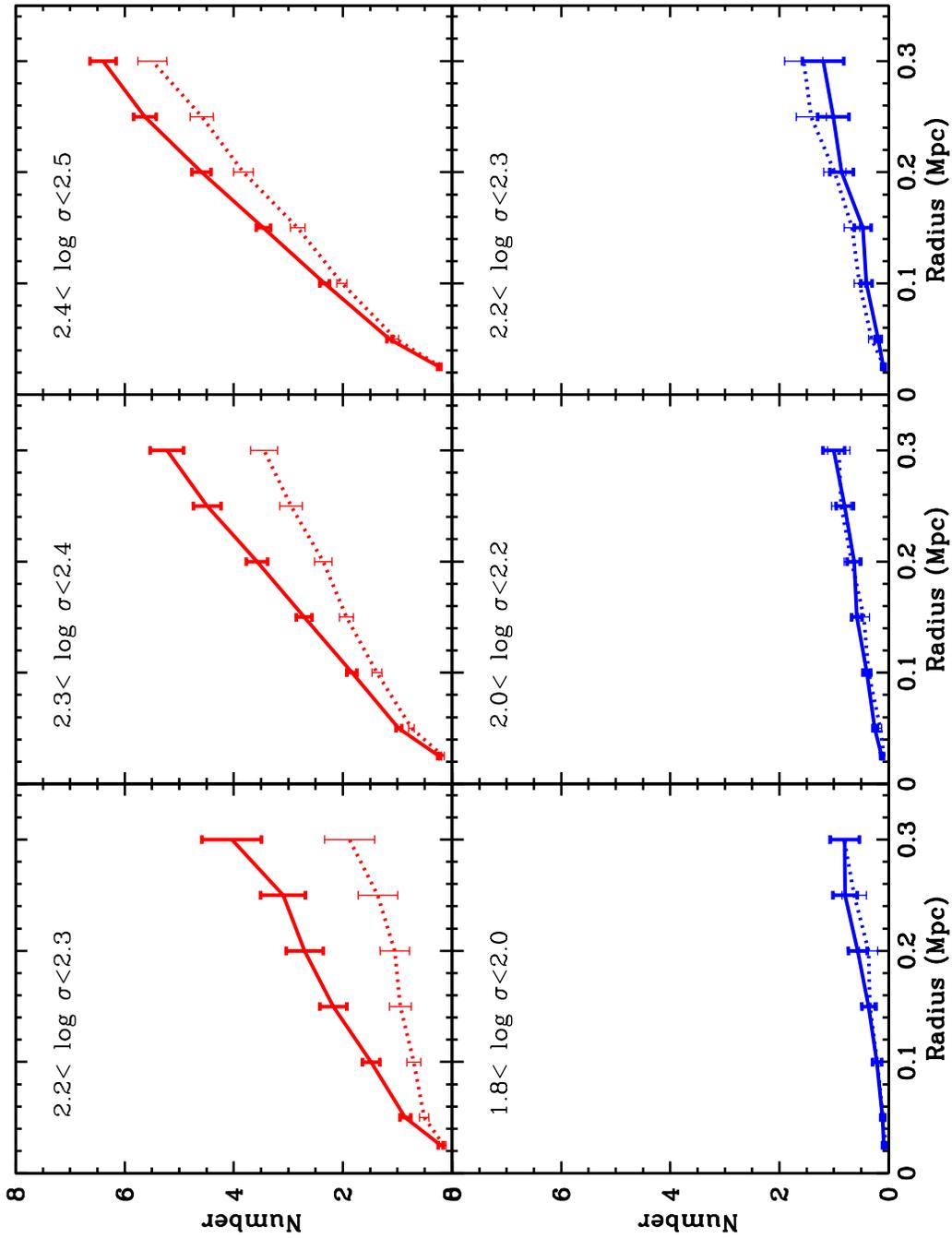}
}
\caption{\label{fig1}
\small
Comparison of the local  environments of radio AGN (upper panels) and  
powerful optical AGN (lower panels) 
with  matched samples of galaxies selected without
regard to their AGN properties (plotted as dotted lines).
We plot the average background-corrected neighbour counts interior to a given radius.}

\end {figure}
\normalsize

The figures discussed above  make it clear that the
conditions that are necessary for triggering 
a radio AGN are quite different to those necessary for triggering
an optical AGN. 
The optical AGN phenomenon requires
that the galaxy have a younger stellar population. The presence of a young
stellar population is presumably linked with a supply of cold gas in the
galaxy that serves as fuel for ongoing star formation and also feeds the black hole.
Powerful optical AGN have very similar structural properties
when compared to typical galaxies of the same velocity dispersion and mass. 
In contrast, the radio AGN phenomenon is not linked to
the presence of young stars and cold gas in the galaxy. 
Radio AGN are triggered in dense environments. This is consistent with the
idea that  radio AGN are  triggered  by {\em hot} rather than cold gas accretion.

What happens when optical and radio AGN occur together, as is the case for  the
majority of powerful radio galaxies at high redshifts? Are the radio and optical 
modes independent or are they able to couple in some way? The fact that radio and
optical line luminosities are strongly correlated in high redshift
radio galaxies would suggest  that the two modes are indeed coupled. In our local sample
of radio AGN with emission lines, a  correlation between emission line and
radio luminosity also appears to be present, but it extends over  a much smaller 
range in luminosity compared to the  higher redshift radio galaxies.   

Another way to look for coupling between the radio and optical phenomena is by using
the matched sample technique. The bottom right panel of Figure 9 shows that
when radio galaxies with emission lines are matched to a sample of radio-quiet
emission line AGN, the distribution of normalized accretion rate (i.e. L[OIII]/M$_{\rm BH}$)
is identical for the two samples. This demonstrates that the presence of the radio
jet does  not influence the optical line emission from the AGN.  

We now  turn the analysis around and ask whether the presence of a strong optical
AGN has any influence on the radio properties of the host galaxy. To do this, we
compare the radio luminosities of the sample of powerful AGN described above 
with a matched sample of galaxies selected without regard to their AGN properties.   
In order to account for the fact that powerful optical AGN have younger
stellar populations and more ongoing star formation, the control sample
is matched in 4000 \AA\ break strength in addition to redshift, stellar mass
and velocity dispersion. 
Our results are shown in Figure 17. We plot the fraction of galaxies in
the two samples as a function of radio luminosity. The first bin includes
galaxies that were not detected in the radio down to a flux limit of 2.5 mJy
(91.6\% of the galaxies in  our sample of strong optical AGN \footnote {It is interesting
that the fraction of nearby  strong optical AGN that are  radio-loud  is very close
to the value of 8\%$\pm$1\% found for SDSS quasars by Ivezi{\'c} et al (2002)} 
 versus 96.7\%
of the galaxies in our set of 10 control samples). Note that we plot
the luminosities of {\em all} galaxies with radio detections independent
of whether they are classified as star-forming or radio AGN. As can be seen,
strong optical AGN have an enhanced probability of being detected in the radio.
The enhanced detection rates persist out to radio luminosities  in excess
of $10^{24}$ W Hz$^{-1}$, where there are no longer any galaxies 
in which the radio emission can be attributed to star formation (see Fig. 1).
We thus conclude that the radio and optical phenomena are not
independent.  
  
\begin{figure}
\centerline{
\epsfxsize=8cm \epsfbox{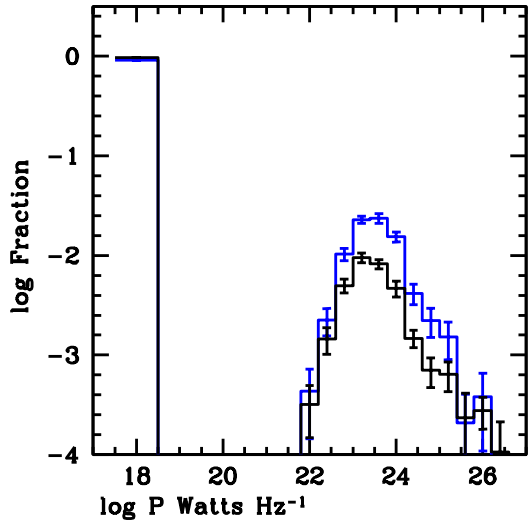}
}
\caption{\label{fig1}
\small
The distribution of radio luminosities for the sample of strong optical
AGN (blue) and the matched samples of galaxies selected without
regard to AGN properties (black). Note that for this plot, the control
samples are matched in 4000 \AA\ break strength in addition to
redshift, stellar mass and stellar velocity dispersion.}
\end {figure}
\normalsize

\section { Is there evidence that feedback from radio jets is  responsible for regulating 
the star formation histories of galaxies?}

In clusters, there is considerable evidence that radio AGN are able to impart 
considerable energy to the  intra-cluster
gas. This may solve the so-called ``cooling flow crisis'' and explain why 
central cluster galaxies have predominantly
old stellar populations and very little ongoing star formation. 
Best et al (2005b,2006,2007) have argued that the
same processes may be responsible for regulating the cooling of gas 
onto the massive galaxy population as a whole.

In principle, the matched sample technique should allow us to investigate
whether the star formation histories of 
radio-loud AGN are any different from radio-quiet galaxies. 
However, in the previous sections we have seen that radio-loud AGN    
are located in denser regions of the Universe  than 
their control galaxies.  It is well known that there is a strong 
correlation  between star formation in a galaxy and its local environment (e.g. Dressler 1980; Balogh et al 1999).
In order to deconvolve the effect of environment and the effect of the 
radio jet, it is therefore necessary
to compare radio-loud AGN  galaxies and control galaxies at a 
fixed value of the local
overdensity parameter.

The results of this exercise are shown in Figure 18. We use the galaxy counts  
within a physical radius
$R< 250$ kpc  as our local overdensity parameter. In order to stack galaxies 
at different redshifts, we
do not count galaxies to a fixed limiting apparent magnitude, but to a magnitude limit of
$m_r +2.5$ mag, where $m_r$ is the $r$-band magnitude of the primary galaxy.  
We plot D$_n$(4000) as a function of $\sigma$ for two different density regimes.
Results for radio loud AGN are plotted in red and results for radio-quiet 
matched galaxies are shown in blue.
The solid line indicates the median value of D$_n$(4000) at 
a given value of $\sigma$, while the
dashed line indicates the lower 25th percentile of the distribution.
For massive galaxies with large values of $\sigma$, there is no discernible 
difference between the two samples.
This holds both in high density and in low density regions. For less massive
galaxies in lower density regions, radio loud AGN are shifted to slightly lower values
of D$_n$(4000)  compared to the control galaxies.

The fact that there is very little difference between the stellar populations of  radio-loud  
and radio-quiet massive galaxies  means that the interplay between heating,  
cooling, and star formation must be in a finely-balanced, self-regulated
state. B05b show that the radio-loud duty-cycle is large in the most massive
galaxies, implying that the radio jet is heating the surrounding gas more or less
continuously.  
We speculate that in low mass galaxies that reside in lower mass  halos, 
heating,cooling and star formation  may not be as   
finely balanced. The duty cycle of the radio-loud phase is much smaller, so
the radio source must have a longer-lasting effect.
Galaxies with small values of the local density parameter 
are typically  located in less massive  dark matter halos than galaxies with large values of
this parameter (see Kauffmann et al 2004). 
The fact that the radio loud
AGN in low density environments  have slightly younger stellar populations may indicate that the radio jet switches on
at the time when gas cooling/accretion and its associated star formation are at a  peak in these systems.
In a recent paper, Croston, Kraft \& Hardcastle (2007) have analyzed shock-heated shells of hot gas 
surrounding a nearby radio galaxy. Their estimate of the amount of energy stored in these shells shows that
shock-heating during the early super-sonic expansion phase can have lasting effects on the 
interstellar medium of the host galaxy.

\begin{figure}
\centerline{
\epsfxsize=15cm \epsfbox{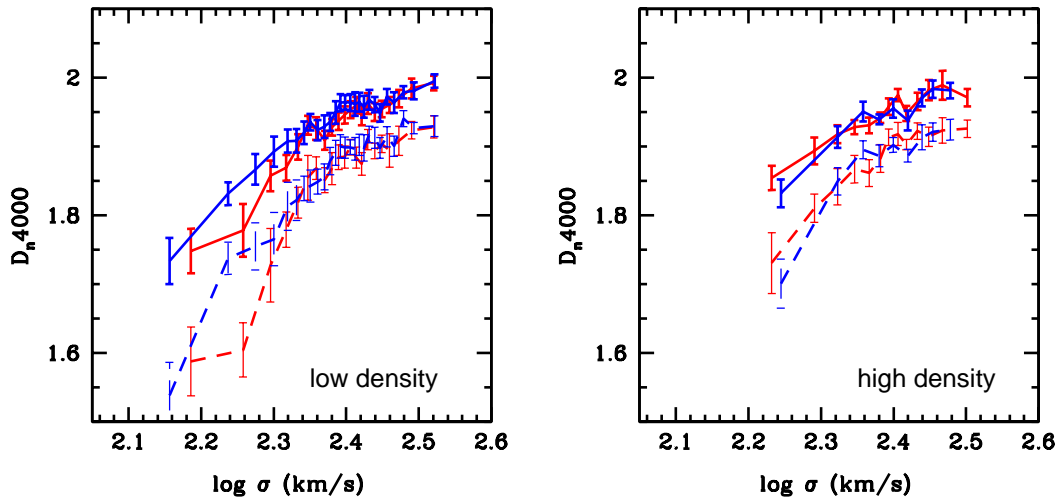}
}
\caption{\label{fig1}
\small
The relation between D$_n$(4000) and velocity dispersion for radio-loud AGN (red) and
a matched sample of  galaxies selected without regard to their
AGN properties (blue). Results are shown for
low density environments (left) and for high density environments (right). The solid line indicates
the median value of D$_n$(4000) at a given value of $\sigma$, while the dashed line
indicates the lower 25th percentile of the distribution.}

\end {figure}
\normalsize

\section {Summary and Discussion}

In the local Universe, the majority of radio-loud AGN are found in massive elliptical
galaxies with weak or absent emission lines and very little ongoing star formation.
Because these galaxies contain very little cold gas, 
the radio jet has been hypothesized to be
powered by accretion from a halo of pressure-supported hot gas 
that surrounds these galaxies (e.g. Allen et al 2006).
Thermal bremsstrahlung from this gas is often  detectable if sufficiently deep  X-ray observations
are available. In the high redshift Universe, radio galaxies are very different.
Almost all powerful  high redshift radio galaxies have strong emission 
lines, and there is ample evidence that the host
galaxies of these systems  are often disturbed and are experiencing a significant
level of ongoing star formation (e.g. Best et al 1998; Pentericci et al 2003; Zirm et al 2003).
 It is thus reasonable to suppose that accretion of  cold
gas may play an important role in these galaxies. 

Our paper focuses on a subset of radio AGN  selected from the Sloan Digital Sky Survey  
that do have emission lines. We explore the hypothesis that these objects 
are local analogs of powerful high-redshift radio galaxies and we ask how they differ
from their nearby radio-quiet counterparts.

In the local Universe, the probability for a  radio AGN to have  emission lines
is a strongly decreasing function of galaxy mass or velocity dispersion. The frequency of
radio AGN with emission lines also rises strongly at radio luminosities greater than
$10^{25}$ W Hz$^{-1}$.  We show that the  emission line and radio luminosities are  
correlated in these objects, but with  substantial scatter.
Interestingly, a significiant component of the scatter can be directly 
attributed to  the velocity dispersion/black hole mass of the host galaxy: 
to reach a given radio power (i.e. a given rate of mechanical energy 
carried by the jet), smaller black holes  require higher [OIII]
luminosities (which we interpret as  higher accretion rates) than bigger black holes.
However, if we scale the emission line and radio luminosities by dividing by $\sigma^4$ (i.e. by
the black hole mass), we find that our sample of RL-AGN define a correlation between
normalized radio power and accretion rate in Eddington units that is independent of black hole mass.
At the present day, EL-RAGN  with high values of $\log P/M_{BH}$   reside in galaxies with smaller
mass black holes than EL-RAGN with low values of  $\log P/M_{BH}$.

Our results lead to a very simple explanation for why very powerful emission line radio galaxies
are much more common at high redshifts. In the local Universe, as a result of
a process that has been dubbed  "down-sizing",  only low mass black holes
are actively growing and accreting at rates close to Eddington  at the
present day. As we have shown, the radio
jets produced by these systems are relatively weak. At high redshifts, massive black
holes are growing and these obejcts will produce much more powerful jets.
Our data  provide some degree of quantitative support for this scenario --
we show that  the relation between emission line and radio luminosity derived 
for  powerful high redshift radio galaxies  matches the correlation 
that we measure for our sample of low-redshift radio galaxies with
large black hole masses  and also
extends it to higher radio power.

In the second part of the paper, we investigate the question of {\it why}  only a small fraction of emission
line AGN become radio loud. We do this by creating matched samples of radio-loud
and radio-quiet AGN and asking whether we can find any systematic differences in their
host galaxy properties and  environments. We find that the main difference between radio-loud
and radio-quiet AGN lies in their environments; radio-loud AGN are found in regions
with local density estimates that are larger by a factor of 2-3. The most plausible  
explanation of this result is that the 
production of luminous radio jets is strongly favoured by  the presence of a
quasi-static, pressure-supported halo of hot gas. With our current
set of observational data,  we  are not able to
tell whether a hydrostatic halo is required in order to {\em form} the jet in the first place,
or whether the halo is merely
required in order dissipate its energy (i.e. the hot gas halo is merely a working surface for
the jet).
Higher resolution radio data (e.g. Gallimore et al 2006) may help distinguish between these two scenarios.

We note that the fact that the radio luminosity of the jet and the strength of the emission
lines are correlated  in radio galaxies with emission lines, 
means that the cold gas  and hot gas phases are not  decoupled. Our  matched sample experiments
show that strong optical AGN have a significantly enhanced probability of hosting radio jets.   
We also find that {\em all} radio AGN with
radio luminosity per unit black hole mass of $10^{17}$ W Hz$^{-1}$ $M_{\odot} ^{-1}$ 
exhibit strong emission lines and evidence of recent star formation, which strongly suggests
that the cold ISM must  play a  role in determining the jet power  in these objects.

The difference between the ``formation" and the ``working surface" hypotheses  
is important for understanding how jets are generated by the black hole,
but less important for understanding the impact of radio AGN on galaxy evolution.
This will mainly depend on how the interplay between cold and hot gas accretion evolves as
a function of redshift.
As has been extensively discussed in
a number of recent papers (Birnboim \& Dekel 2003; Keres et al 2005; Croton et al 2006),
the transition between the accretion of cold gas along  filaments and the accretion of hot gas
from  a quasi-static, pressure-supported halo always occurs at a dark matter halo mass of
$\sim 10^{12} M_{\odot}$. If optical AGN are fuelled primarily by cold
gas accretion and radio AGN by hot gas accretion, we would expect radio AGN to be
located in significantly more massive halos than optical AGN at all redshifts.  
Our local density estimator, which 
counts the  number of neighbours within a few hundred
kiloparsecs of the galaxy, is expected to scale quite closely with the mass of
the surrounding halo (Kauffmann et al 2004). In future work, we will
explore  whether our observational results are in quantitative agreement with
a model in which radio galaxies only reside in halos greater than this minimum mass.   

In this scenario,  radio-loud AGN with strong emission lines  occupy a
restricted region of parameter space; we would expect them to be  located in
environments where both cold and hot gas accretion can occur {\em simultaneously}.
In the low redshift Universe, cold gas is only present in low mass galaxies 
that are usually located  in low mass
halos. The majority of massive galaxies have very little cold gas and a large fraction are located
in halos  considerably more massive than the transition mass of $10^{12} M_{\odot}$.  
This explains why   radio 
galaxies with emission lines  are rare in  our low redshift sample  and why they 
are restricted to galaxies with low values of $\sigma$ located
in denser-than-average environments.  At high redshifts, however,  the situation
will be quite different.  Massive galaxies will be located in considerably smaller halos and 
they may well still be accreting cold  gas and thus be capable of
hosting powerful radio AGN with emission lines. This will be the subject of future work.

Finally, we have explored whether radio-loud AGN have systematically different star formation
histories compared to matched samples of radio-quiet galaxies. This comparison is interesting, 
because it sheds light on the manner in which radio jets regulate the growth of galaxies through star
formation. Because star formation rate and local galaxy density 
are known to be correlated, and because  radio AGN are  found
in denser-than-average regions of the Universe, we carry out our comparison between 
radio-loud and radio-quiet objects at a fixed local density.

In the most massive galaxies and in the densest environments, the duty cycle of the radio
AGN phenomenon is very high (B05b show that the fraction of radio AGN reaches values
close to 50\% for the most massive galaxies on the local Universe). 
It is perhaps not too surprising,  therefore, that 
for these systems the age of the stellar population does not depend on  whether the 
galaxy is radio-loud or radio-quiet. The radio jet is heating the gas more or
less continuously and star formation rates are maintained at very low levels.
It is only for low mass galaxies located in low-density environments 
(i.e. in low mass halos) that significant differences between the
radio-loud AGN and control  samples of normal galaxies begin to emerge. We find that in these systems
radio jets are preferentially located in galaxies with systematically younger 
stellar populations. This means that in low mass halos,  the radio jet is dumping its energy in the vicinity
of those objects that are currently experiencing higher-than-average rates of star
formation (see Figure 18).                                                      

In the local Universe,  radio AGN in young, star-bursting galaxies are rare and 
are not intrinsically very powerful because they are powered by relatively 
low mass black holes. However, at higher redshifts massive black holes  are alive
and awake and the effects of the much more powerful radio jet may be considerably more dramatic (Nesvadba et al 2006).
As more and more data on high redshift galaxies is accumulated, it should become
increasingly possible to apply the lessons learned from studying galaxies
and AGN at  low redshifts in order to
figure out which physical  processes dominate as a function of  cosmic epoch.

\vspace{0.3cm}
\large
\noindent
{\bf Acknowledgements}\\
\normalsize

\noindent
We thank Roderik Overzier for stimulating, useful and amusing discussions.

\noindent
Funding for the creation and distribution of the SDSS Archive has been
provided by the Alfred P. Sloan Foundation, the Participating Institutions,
the National Aeronautics and Space Administration,
the National Science Foundation, the U.S. Department of Energy,
the Japanese Monbukagakusho, and the Max Planck Society.
The SDSS Web site is http://www.sdss.org/.
The SDSS is managed by the Astrophysical Research Consortium (ARC)
for the Participating Institutions. The Participating Institutions
are The University of Chicago, Fermilab, the Institute for Advanced Study,
the Japan Participation Group, The Johns Hopkins University,
the Korean Scientist Group, Los Alamos National Laboratory,
the Max-Planck-Institute for Astronomy (MPIA),
the Max-Planck-Institute for Astrophysics (MPA),
New Mexico State University, University of Pittsburgh,
University of Portsmouth, Princeton University,
the United States Naval Observatory, and the University of Washington.

\pagebreak
\Large
\begin {center} {\bf References} \\
\end {center}
\normalsize
\parindent -7mm
\parskip 3mm

Allen S.~W., Dunn R.~J.~H., Fabian A.~C., Taylor G.~B., Reynolds C.~S., 
2006, MNRAS, 372, 21 

Baldwin J.~A., Phillips M.~M., 
Terlevich R., 1981, PASP, 93, 5 (BPT) 

Balogh M.~L., Morris S.~L., Yee H.~K.~C., Carlberg R.~G., Ellingson E., 
1999, ApJ, 527, 54 

Barthel, P.~D., 1989, ApJ, 336, 606

Baum S.~A., Zirbel E.~L., O'Dea C.~P., 1995, ApJ, 451, 88 

Becker R.~H., White R.~L., Helfand 
D.~J., 1995, ApJ, 450, 559 

Best, P.~N., Longair,
M.~S., \& R\"ottgering, H.~J.~A., 1998, MNRAS, 295, 549

Best P.~N., Kauffmann G., Heckman T.~M., Brinchmann J., Charlot S., 
Ivezi{\'c} {\v Z}., White S.~D.~M., 2005a, MNRAS, 362, 25 (B05a)

Best P.~N., Kauffmann G., Heckman T.~M., Ivezi{\'c} {\v Z}., 2005b, MNRAS, 
362, 9 (B05b)

Best P.~N., Kaiser C.~R., Heckman T.~M., Kauffmann G., 2006, MNRAS, 368, 
L67 

Best P.~N., von der Linden A., Kauffmann G., Heckman T.~M., Kaiser C.~R., 
2007, MNRAS, in press

Birnboim Y., Dekel A., 2003, MNRAS, 345, 349 

Birzan L., Rafferty D.~A., McNamara B.~R., Wise M.~W., Nulsen P.~E.~J., 2004, 
ApJ,607,800

Boehringer H., Voges W.,Fabian A.~C., Edge A.~C., Neumann D.~M., 1993, MNRAS,
264, L25

Bower R.~G., Benson A.~J., Malbon R., Helly J.~C., Frenk C.~S., Baugh 
C.~M., Cole S., Lacey C.~G., 2006, MNRAS, 370, 645

Brinchmann J., Charlot S., White S.~D.~M., 
Tremonti C., Kauffmann G., Heckman T., Brinkmann J., 2004, MNRAS, 351, 1151 

Bruzual G., Charlot S., 2003, MNRAS, 
344, 1000 

Charlot S., Fall S.~M., 2000, ApJ, 539, 718

Churazov E., Brueggen M., Kaiser C.~R., Boehringer H., Forman W., 2001, ApJ,
554, 261

Clewley L., Jarvis M.~J., 2004, MNRAS, 
352, 909 

Condon J.~J., Cotton W.~D., Greisen E.~W., Yin Q.~F., Perley R.~A., Taylor 
G.~B., Broderick J.~J., 1998, AJ, 115, 1693 

Croston J.~H., Kraft R.~P., 
Hardcastle M.~J., 2007, ApJ, 660, 191 

Croton D.~J., et al., 2006, MNRAS, 365, 11 

Dressler A., 1980, ApJ, 236, 531

Dunlop J.~S., Peacock J.~A., 1990, MNRAS, 247, 19 

Di Matteo T., Springel V., Hernquist 
L., 2005, Natur, 433, 604 

Ellingson E., Yee H.~K.~C., Green R.~F., 
1991, ApJ, 371, 49 

Fabian A.~C., et al., 2000, MNRAS, 318, L65                          

Fabian A.~C., Sanders J.~S., Allen S.~W., Crawford C.~S., Iwasawa K.,
Johnstone, R.~M., Schmidt R.~W., Taylor G.~B., 2003, MNRAS, 344, L43 

Fabian A.~C., Sanders J.~S., Taylor G.~B., Allen S.~W., 2005, MNRAS, 360, L20

Fanaroff B.~L., Riley J.~M., 1974, MNRAS, 167, 31P

Forman, W., et al., 2005, ApJ, 635, 894                           

Gallimore J.~F., Axon D.~J., O'Dea C.~P., 
Baum S.~A., Pedlar A., 2006, AJ, 132, 546 

Govoni F., Falomo R., Fasano G., Scarpa R., 2000, A\&A, 353, 507 

Hardcastle M.~J., Evans D.~A., Croston 
J.~H., 2006, MNRAS, 370, 1893 

Hardcastle M.~J., Evans D.~A., Croston 
J.~H., 2007, MNRAS, 376, 1849 

Heckman T.~M., Kauffmann G., Brinchmann 
J., Charlot S., Tremonti C., White S.~D.~M., 2004, ApJ, 613, 109 

Hine R.~G., Longair M.~S., 1979, 
MNRAS, 188, 111 

Hopkins P. F., Hernquist L., Cox, T.J., Di Matteo T., Martini P.,
Robertson, B., Springel, V., 2005, ApJ, 630, 705

Ivezi{\'c} {\v Z}., et al., 2002, AJ, 124, 
2364 

Kaiser C.~R., Dennett-Thorpe J., 
Alexander P., 1997, MNRAS, 292, 723 

Kauffmann G., et al., 2003a, MNRAS, 341, 33 

Kauffmann G., et al., 2003b, MNRAS, 346, 1055 

Kauffmann G., White S.~D.~M., Heckman 
T.~M., M{\'e}nard B., Brinchmann J., Charlot S., Tremonti C., Brinkmann J., 
2004, MNRAS, 353, 713 

Kere{\v s} D., Katz N., Weinberg D.~H., 
Dav{\'e} R., 2005, MNRAS, 363, 2 

Laing R.~A., Riley J.~M., Longair 
M.~S., 1983, MNRAS, 204, 151 

Laing R.~A., Jenkins C.~R., Wall J.~V., Unger S.~W., 1994, ASPC, 54, 201 

Ledlow M.~J., Owen F.~N., 1995, AJ, 109, 853

Li C., Kauffmann G., Wang L., White S.~D.~M., Heckman T.~M., Jing Y.~P., 2006, 
MNRAS, 373, 457 

McCarthy P.~J., 1993, ARA\&A, 31, 639 

McLure, R.~J., Dunlop, J.~S., 2001, MNRAS, 321, 515

Nesvadba N.~P.~H., et al., 2006, ApJ, 650 639, 

Pentericci, L.,
R{\"o}ttgering, H.~J.~A., Miley, G.~K., McCarthy, P., Spinrad, H., van
Breugel, W.~J.~M., \& Macchetto, F.,1999, A\&A , 341, 329

Sijacki D., Springel V., Di Matteo T., 
Hernquist L., 2007, arXiv, 705, arXiv:0705.2238 

Smith E.~P., Heckman T.~M., 1989, ApJ, 
341, 658 

Tadhunter C., Dickson R., Morganti R., 
Robinson T.~G., Wills K., Villar-Martin M., Hughes M., 2002, MNRAS, 330, 
977 

Tadhunter C., Robinson T.~G., Gonz{\'a}lez 
Delgado R.~M., Wills K., Morganti R., 2005, MNRAS, 356, 480 

Tremaine S., et al., 2002, ApJ, 574, 740 

Tremonti C.~A., et al., 2004, ApJ, 613, 898

Wild V., Kauffmann G., Heckman T., Charlot S., Lemson G., Brinchmann J., 
Reichard T., Pasquali A., 2007, arXiv, 706, arXiv:0706.3113 

Worthey G., Ottaviani D.~L., 1997, 
ApJS, 111, 377 

Yee, H.~K.~C., Green, R.~F., 1984, ApJ, 280, 79

Yun M.~S., Reddy N.~A., Condon J.~J., 
2001, ApJ, 554, 803 

Zirbel E.~L., Baum S.~A., 1995, ApJ, 448, 521 (ZB) 

Zirm, A.~W., Dickinson,
M., \& Dey, A., 2003, ApJ, 585, 90

\end{document}